\documentclass[a4paper,12pt]{article}

\usepackage{geometry}
\usepackage{amsmath}
\usepackage{amssymb}
\usepackage{graphicx} 
\usepackage{epstopdf} 
\usepackage[font=footnotesize,skip=10pt,justification=centering,belowskip=-10pt,aboveskip=10pt]{caption}
\usepackage{subcaption}
\usepackage{multirow} 
\usepackage{textcomp} 
\usepackage{stmaryrd}
\usepackage{makeidx}
\usepackage{epsfig}
\usepackage{float}
\usepackage{wasysym} 
\usepackage{authblk} 
\usepackage{algorithm2e}
\usepackage{dirtytalk}
\usepackage{xcolor}

\usepackage[english]{babel}
\usepackage{hyperref}
\hypersetup{pdftex, colorlinks=true, linkcolor=blue, citecolor=blue, filecolor=blue, urlcolor=blue, pdftitle=, pdfauthor=, pdfsubject=, pdfkeywords=}

\newcounter{Eqn}

\newcounter{Refs}

\title{\bf{Martensite decomposition kinetics in additively manufactured Ti-6Al-4V alloy: in-situ characterisation and phase-field modelling}}

\author[a,b,c,*]{A.D.~Boccardo}
\author[d]{Z.~Zou}
\author[d]{M.~Simonelli}
\author[a,b]{M.~Tong}
\author[e,c]{J.~Segurado}
\author[a,b]{S.B.~Leen}
\author[c,*]{D.~Tourret}

\affil[a]{\small{Mechanical Engineering, School of
    Engineering, College of Science and Engineering, University of Galway,
    University Road, Galway H91 HX31, Ireland.}}

\affil[b]{\small{I-Form Advanced Manufacturing Research Centre,
    University of Galway, University Road, Galway H91 HX31, Ireland.}}

\affil[c]{\small{IMDEA Materials Institute, C/ Eric Kandel 2,
    28906, Getafe, Madrid, Spain. $^{*}$email
    corresponding authors: adrian.boccardo@imdea.org, damien.tourret@imdea.org}}

\affil[d]{\small{Centre for Additive Manufacturing, Faculty of
    Engineering, University of Nottingham, Advanced Manufacturing
    Building, Jubilee Campus, United Kingdom.}}

\affil[e]{\small{Universidad Polit\'ecnica de Madrid, Department
    of Materials Science, E.T.S.I. Caminos, C/ Profesor Aranguren 3,
    28040, Madrid, Spain. 
    }}

\date{ }

\begin{document}
\maketitle

\begin{abstract}

  Additive manufacturing of Ti-6Al-4V alloy via laser powder-bed fusion leads to non-equilibrium $\alpha'$
  martensitic microstructures, with high strength but poor ductility and toughness. These properties may be
  modified by heat treatments, whereby the $\alpha'$ phase decomposes into equilibrium $\alpha+\beta$
  structures, while possibly conserving microstructural features and length scales of the $\alpha'$ lath structure.
  Here, we combine experimental and computational methods to explore the kinetics of martensite
  decomposition. Experiments rely on \textit{in-situ} characterisation (electron
  microscopy and diffraction) during multi-step heat treatment from 400$^{\circ}$C up to the alloy $\beta$-transus
  temperature (995$^{\circ}$C). Computational simulations rely on an experimentally-informed computationally-efficient phase-field model. 
  Experiments confirmed that as-built microstructures were fully composed of martensitic $\alpha'$ laths. During martensite
  decomposition, nucleation of the $\beta$ phase occurs primarily along $\alpha'$ lath boundaries, with traces of
  $\beta$ nucleation along crystalline defects. Phase-field results, using electron backscatter
  diffraction maps of as-built microstructures as initial conditions, are compared directly with \textit{in-situ}
  characterisation data. Experiments and simulations confirmed that, while full decomposition into stable $\alpha+\beta$ phases
  may be complete at 650$^{\circ}$C provided sufficient annealing time, visible morphological evolution of the microstructure was only observed for $T\geq\,$700$^{\circ}$C, without modification of the prior-$\beta$ grain structure.
\end{abstract}

\vspace{1pc}

%
\section{Introduction}
%

\begin{figure}[b!]
  \begin{center}
    \includegraphics[width=3.5in]{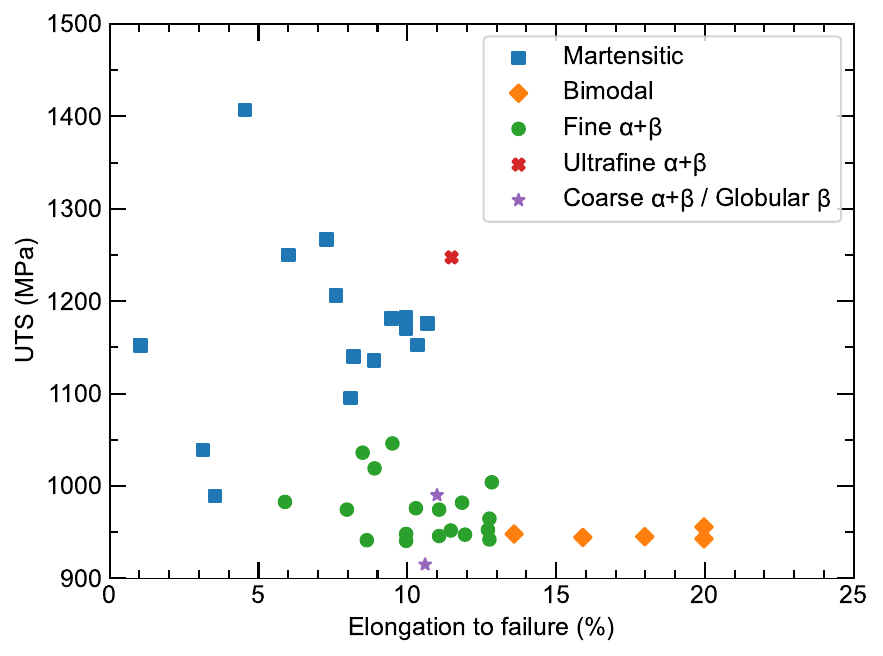}
    \caption{Ultimate tensile strength versus elongation to failure of Ti-6Al-4V samples
      built by L-PBF, with different types of microstructures, either martensitic (as built)
      or after heat treatment. Data compiled from Refs \cite{bkm:RefVandenbrouke2007-6,bkm:RefFacchini2010-7,bkm:RefVrancken2012-8,bkm:RefMurr2009-9,bkm:RefVilaro2011-10,bkm:RefXu2015-11,bkm:RefXu2016-12,bkm:RefTerHaar2018-13,bkm:RefKaschel2020a-14}.}
    \label{bkm:FigMechProp}
  \end{center}
\end{figure}

Titanium (Ti) alloys, such as Ti-6Al-4V (weight \%), play a central role in modern structural applications, in
particular in aeronautics and biomedical applications \cite{bkm:RefLeyens2003-1,bkm:RefBanerjee2013-2}. 
Additive manufacturing (AM) technologies, such as laser powder bed fusion (L-PBF) have demonstrated a great potential for the manufacture of high-performance Ti alloys, thanks to their ability to produce near-net shape components of complex geometry with
minimal material waste \cite{bkm:RefFrasier2014-3,bkm:refHerzog2016-4,bkm:RefBecker2021-5}. L-PBF printed Ti-6Al-4V parts -- with or
without post-AM heat treatment -- exhibit a broad range of microstructures, and a commensurate range of mechanical properties (Figure~\ref{bkm:FigMechProp}). 

Equilibrium phases in Ti-6Al-4V consist of a hexagonal close-packed (hcp) $\alpha$ phase and
a body-centred cubic (bcc) $\beta$ phase, both of which are common in wrought or cast alloys (i.e. 
with low-to-moderate cooling rates) \cite{bkm:RefWelsch1998-15,bkm:RefElmer2005-16}. 
Nonequilibrium phases include a hcp $\alpha'$ martensite obtained by rapid cooling \cite{bkm:RefXu2015-11,bkm:RefXu2016-12,bkm:RefWelsch1998-15} and an orthorhombic soft $\alpha''$ martensite \cite{bkm:RefKolichev1999-17,bkm:RefWilliams1970-18,bkm:RefFroes2015-19},
which may form in localised regions with high vanadium concentrations (typically 9 to 13 wt\%) \cite{bkm:RefBoyer1994-20}.

While minor amounts of secondary phases have been reported \cite{bkm:RefZafari2018-21,bkm:RefHaubrich2019-22,bkm:RefMurr2009-9,bkm:RefThijs2010-23}, as-printed L-PBF Ti-6Al-4V most often exhibits a fully acicular martensitic $\alpha'$ microstructure \cite{bkm:RefMurr2009-9,bkm:RefVilaro2011-10,bkm:RefXu2015-11,bkm:RefThijs2010-23,bkm:RefSercombe2008-24,bkm:RefSong2012-25,bkm:RefWielewski2012-26,bkm:RefSimonelli2014-27,bkm:RefYang2016-28,bkm:RefWu2016-29,bkm:RefKasperovich2015-30,bkm:RefTan2016-31,bkm:RefKrakhmalev2016-32,bkm:RefHuang2016-33,bkm:RefBarrioberoVila2017-34,bkm:RefCao2018-35,bkm:RefZhang2018-36}.
The $\alpha'$ lath structure forms within prior-$\beta$ grains, following Burgers crystallographic
relationships, thus allowing the reconstruction of parent $\beta$ grains \cite{bkm:RefWielewski2012-26,bkm:RefSimonelli2014-27,bkm:RefHumbert1995-37,bkm:RefHumbert1996-38,bkm:RefGlavicic2003-39,bkm:RefGlavicic2003b-40}.
Primary $\beta$ grains typically exhibit elongated shapes with strong $\langle100\rangle$ texture along
their growth direction, but the resulting $\alpha'$ martensite has a weak texture due to the presence of many
martensitic variants \cite{bkm:RefSimonelli2014-27,bkm:RefFormanoir2016-41,bkm:RefKarami2020-42,bkm:RefPantawane2021-43}.

Martensitic microstructures offer high strength but low ductility, due to the hcp nature and slightly distorted
lattice of the supersaturated $\alpha'$ phase, making it undesirable for applications requiring a high elongation to failure \cite{bkm:RefFacchini2010-7,bkm:RefVrancken2012-8,bkm:RefWelsch1998-15,bkm:RefDonachie2000-44}.
Since dual $(\alpha+\beta)$ microstructures exhibit excellent mechanical properties (Fig.~\ref{bkm:FigMechProp}), post-printing heat treatment is often applied to improve mechanical properties (e.g. ductility) \cite{bkm:RefVilaro2011-10,bkm:RefKaschel2021-45, liu2022effect, xiao2022mechanism, dhekne2023micro, li2023optimizing}. 
Interestingly, beyond mechanical properties, post-printing heat treatment and the resulting
microstructure evolution were also found to substantially improve corrosion resistance \cite{bkm:RefLiu2019-58,bkm:RefHemmasian2019-59,bkm:RefZhang2021electrochem-60}
and biocompatibility (e.g. hydrophilicity and surface roughness promoting early cell attachment, proliferation and
osseointegration, as well as good cytocompatibility) \cite{bkm:RefWang2016-61}.

Heat treatments below the $\beta$-transus temperature ($\approx995^{\circ}$C) and above 400$^{\circ}$C allow for the martensite
decomposition ($\alpha' \to \alpha+\beta$), while conserving the lamellar features and length scales of the $\alpha'$ lath structure
\cite{bkm:RefVrancken2012-8,bkm:RefVilaro2011-10,bkm:RefWu2016-29,bkm:RefZhang2018-36,bkm:RefLeuders2014-46,bkm:RefKasperovich2015-47,bkm:RefGalarraga2016-48,bkm:RefGalarraga2017-49,bkm:RefBaker2017-50,bkm:RefZou2020-51}.
Below 400$^{\circ}$C, stress relaxation of the crystal lattice occurs without apparent phase transformation \cite{bkm:RefKaschel2021-45}.
Above 400$^{\circ}$C, martensite decomposition typically occurs with negligible influence of heating \cite{bkm:RefKaschel2021-45} or cooling \cite{bkm:RefKasperovich2015-47,bkm:RefWycisk2014-52} rates. 
A broad range of experimental studies have confirmed that the kinetics of martensite decomposition is limited by the diffusion of excess vanadium from the $\alpha'$ martensite
\cite{bkm:RefTerHaar2018-13,bkm:RefHaubrich2019-22,bkm:RefTan2016-31,bkm:RefBarrioberoVila2017-34,bkm:RefZeng2005-62,bkm:RefAlBermani2010-63,bkm:RefSallicaLeva2016-64,bkm:RefKazantseva2018-65,bkm:RefGupta2016-66,bkm:RefKaschel2020b-67}.
While the enrichment in $\beta$-stabilisers at one- and two-dimensional lattice defects may promote $\beta$-phase nucleation along such defects \cite{bkm:RefHaubrich2019-22,bkm:RefZou2020-51}, the most common $\beta$ phase nucleation sites were reported along V segregated grain boundaries between $\alpha'$ laths \cite{bkm:RefTan2016-31,bkm:RefChao2014-68,bkm:RefGhosh2022-69}.

Traditional heat treatments utilised on L-PBF Ti-6Al-4V were designed for significantly different thermo-mechanical
processes (and hence microstructures) \cite{bkm:RefZhang2018-36,bkm:RefBaker2017-50}, and may thus be sub-optimal, or even not appropriate, for additively manufactured Ti-6Al-4V. As a result, new heat treatment processes need to be designed and tailored to L-PBF Ti-6Al-4V \cite{bkm:RefZou2020-51,bkm:RefSabban2019-70}.
Here, we argue that the exploration and optimisation of novel heat treatments suited to L-PBF Ti-6Al-4V can be greatly accelerated by the development of state-of-the-art experimentally-informed simulations of microstructure evolution, supported by advanced \textit{in-situ} characterisation techniques.

The rapid advance of \textit{in-situ} characterisation of metals has provided key insight into
microstructure formation and evolution \cite{bkm:RefElmer2005-16,bkm:RefZou2020-51,bkm:RefKaschel2020b-67,bkm:RefChen2011-71,bkm:RefCalta2020-72}.
Still, experimental techniques suffer limitations, in particular when multiple characteristics should ideally be tracked simultaneously.
For instance, simultaneous monitoring of crystal structure/orientation jointly with chemical composition remains challenging. In that
context, the use of computational models can provide a critical
support to experimental-based interpretation, i.e. to \say{fill in the blanks} -- e.g. complementing measurements limited by the finite response time and spatial accuracy of detectors, in particular at high temperatures -- and reach a deeper understanding of microstructural evolution.

From the modelling perspective, martensite decomposition in Ti-6Al-4V has been studied using classical mean-field Avrami-based models \cite{bkm:RefGilMur1996-73,bkm:RefMurgau2012-74,bkm:RefYang2021-75,bkm:RefBAYKASOGLU,bkm:met8080633,bkm:SUN}. 
While mean-field models provide a fast and convenient tool to assess the evolution
of phase fractions at different temperatures, they lack information on microstructure morphology and chemical composition, which have a significant effect on mechanical properties, even when phase fractions are equivalent (Figure~\ref{bkm:FigMechProp}). 
Microstructure-aware models, such as phase-field (PF) approaches, while more computationally expensive, offer a more detailed description of the evolution of microstructural features, and of their coupling with chemical composition fields \cite{chen2002phase, bkm:RefShi2016-77,bkm:RefJi2018-78,bkm:RefTourret2022-76}. 
Phase-field models have been proposed to simulate the evolution of Ti-6Al-4V microstructures (phases and composition), for instance for the formation of $\beta$ microstructure and subsequent precipitation and dissolution of precipitates during repeated thermal cycles \cite{shi2019integrated}, or for the evolution of an \{$\alpha+\beta$\} microstructure subjected to annealing treatment \cite{bkm:RefHuang2019-79,bkm:RefAhluwalia2020-80,bkm:RefZhang2021-81}. 
Moreover, as mentioned in Ref. \cite{bkm:RefJi2018-78}, the predictive capability of PF models depends strongly on the accuracy of
the input materials data, in particular temperature-dependent thermodynamic and kinetic parameters, making direct comparison to
experiments absolutely essential. 
Yet, to the best of our knowledge, PF models focused on the decomposition of $\alpha'$ martensite into $\alpha$ and $\beta$ phases remain lacking.

Here, we propose an experimentally informed phase-field model for martensite
decomposition in additively manufactured Ti-6Al-4V alloy. 
The underlying goal is to develop a digital tool to study the evolution of Ti-6Al-4V microstructure, so as to ultimately guide the design of novel heat treatments for optimal properties in additively manufactured Ti-6Al-4V. 
The proposed PF model goes beyond existing mean-field models for martensite decomposition by predicting the coupled spatial evolution of phase and solute in the heterogeneous microstructure.
The model parameters are calibrated and validated using experimental data from original \textit{in-situ} annealing experiments on L-PBF manufactured Ti-6Al-4V. 
In order to address the computational cost of PF simulations, we solve PF equations by means of an original spectral Fourier-based method computationally parallelised on graphics processing units (GPUs), improving upon a method recently introduced in \cite{bkm:RefBoccardo2023-91}. 
The resulting model is robust and efficient, it considers heterogeneous diffusivities in the different phases, and it allows for the use of initial $\alpha'$ martensite microstructures directly obtained from electron backscattered diffraction (EBSD) maps. 
We use the PF model to simulate the microstructure evolution under stepwise heating treatment and directly compare volume fractions and microstructural morphologies against our \textit{in-situ} high-temperature EBSD data.

%
\section{Materials and methods}
%

%
\subsection{Experiments}
Samples used in this study were made of Ti-6Al-4V (grade 23) specimens produced by L-PBF on an EOSINT M290 printer using
proprietary optimised process parameters \cite{bkm:RefZou2021-82}. Microstructural constituents of the as-built material were examined
using X-ray Diffraction (XRD) with a Bruker D8 ADVANCE device in conjunction with the DAVINCI XRD system. The XRD pattern
was scanned via a \text{Cu-K$\alpha$1} X-ray source using a step size of 0.02$^{\circ}$ and a time step of 1.5~s.

\textit{In-situ} microstructural observations were carried out on the frontal plane ($xz$-surface) of the specimens using a JEOL
7100F FEG-SEM equipped with a heating Scanning Electron Microscope (SEM) stage (Murano in-situ stages, Gatan), using a focus distance of
10~mm, emission energy of 15~kV, probe current of 8~\textmu A, and step size of 0.05~\textmu m. Flat specimen of 7~mm$\times$7~mm
and 1.5~mm thick was mirror polished and then mounted on the heating stage via high-temperature carbon paste. The heating temperature was measured via a
thermocouple attached to the bottom of the specimen.
Such settings allow controlled heating of specimens up to 980$^{\circ}$C, with simultaneous imaging
and EBSD. Where needed, the area fraction of the $\beta$ phase was quantified either using the EBSD data using HKL-Channel~5{\texttrademark}, or using
the SEM images using FIJI (ImageJ) via a Random Forest machine learning algorithm -- as reported in detail elsewhere
\cite{bkm:RefMiyazaki2019-83}.

\begin{figure}[h!]
  \begin{center}
    \includegraphics[width=3.5in]{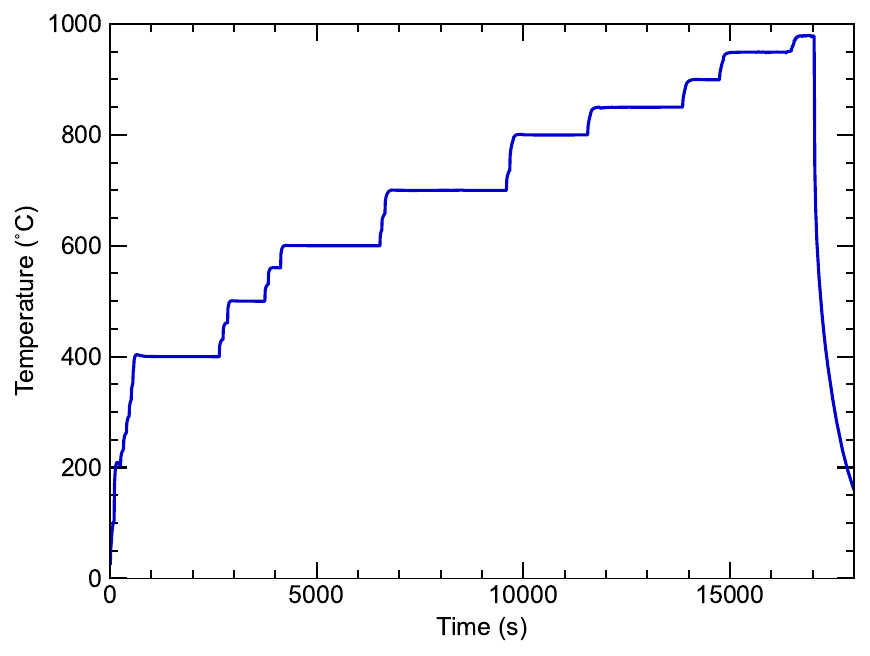}
    \caption{Temperature-time profile employed in the annealing process.}
    \label{bkm:FigTempVsTime}
  \end{center}
\end{figure}

A multi-step thermal cycle, shown in
Figure~\ref{bkm:FigTempVsTime}, was applied to the sample during the \textit{in-situ} characterisation, in order
to study the microstructure evolution at different annealing temperatures during the process.

\subsection{Modelling}
\subsubsection{Microstructure representation}
In order to simulate the microstructure evolution during the multi-step annealing process, we used a numerical model
based on the phase-field method. The model considers the bcc phase $\beta$, and hcp phases $\alpha$ (equilibrium) and
$\alpha'$ (martensite). Since the martensite ($\alpha'$) phase may essentially be considered as a distorted
equilibrium ($\alpha$) phase supersaturated in solute, both $\alpha$ and
$\alpha'$ phases are actually represented as the same phase, and they are
differentiated from each other by their solute concentration. Hence, the phase field (order parameter), $\eta$,
is equal to zero in the $\beta$ phase and one in $\alpha$ and/or $\alpha'$ phases. Different orientations
(variants) of the $\alpha'$ (or $\alpha$) phase are represented by different phase fields $\eta_{i}$ as
commonly done in so-called multi-phase-field \cite{bkm:RefSteinbach1996-84,bkm:RefEiken2006-85,bkm:RefSteinbach2006-86}
or multi-order-parameter \cite{bkm:RefOforiOpoku2010-87} approaches. Specifically, we consider
$p$ orientations of hcp phases. With  $1\leq i \leq p$, only one $\eta_{i}$ reaches unity at a given point,
corresponding to a hcp phase of orientation index  $i$, while all $\eta_{i}$ are equal to zero at
$\beta$ phase locations. 
Conceptually, the model could be extended to different to multiple $\beta$ grains.
However, here we consider only one $\beta$ orientation, since we focus on the decomposition kinetics of a single prior-$\beta$ grain, shown by experimental data (Section~\ref{bkm:Ref122788851}) to result in the same crystal orientation as the initial (prior) $\beta$ grain.

\begin{figure}[h!]
  \begin{center}
    \includegraphics[width=2.5in]{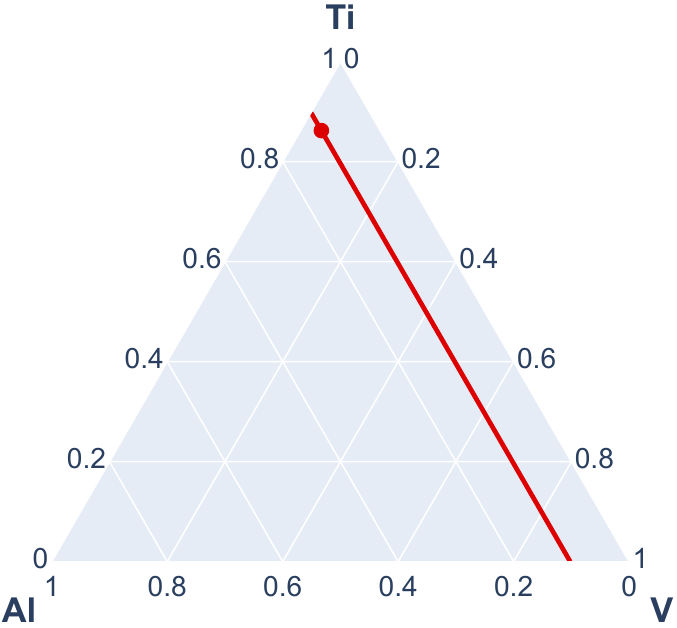}
    \caption{Pseudo-binary TiAl-V isopleth represented as a red line within the ternary Ti-Al-V system. For the sake of consistency with the model, units are in mole fraction. The nominal Ti-6Al-4V (wt\%) alloy composition is marked by a circle at $c_{\rm Al}=0.102$ and $c=c_{\rm V}=0.036$.
    }
    \label{bkm:FigPseudoBinary}
  \end{center}
\end{figure}

It is recognised that the kinetics of martensite decomposition in Ti-6Al-4V alloy is limited by the diffusion of V solute
\cite{bkm:RefHaubrich2019-22,bkm:RefTan2016-31,bkm:RefBarrioberoVila2017-34,bkm:RefAlBermani2010-63,bkm:RefSallicaLeva2016-64,bkm:RefKazantseva2018-65}.
Therefore, for the sake of simplicity, solute diffusion and chemical free energies were calculated considering a pseudo-binary TiAl-V system,
defined along the isopleth at $c_{\rm Al}=0.102$ (i.e. 10.2\,at.\%Al), as illustrated by the red line in
Figure~\ref{bkm:FigPseudoBinary}, and V is the sole considered solute element ($c\equiv c_\text{V}$) in the model. Note that, unless mentioned otherwise, all concentrations in the article are expressed in mole fraction.

As we specifically aim at modelling the martensite decomposition transformation ($\alpha' \to \alpha+\beta$)
following L-PBF, the initial (as-built) microstructure is assumed to be fully martensitic ($\alpha'$) with grains of
different crystal orientations, dictated by the orientation relations with prior-$\beta$ grains. The nucleation
of $\beta$ grains was reported to occur along grain boundaries of the $\alpha'$ phase needles due to the local
enrichment in vanadium \cite{bkm:RefTan2016-31,bkm:RefChao2014-68,bkm:RefGhosh2022-69}. Therefore, we chose to take
advantage of the spontaneous nucleation of $\beta$ at $\alpha'$ lath boundaries due to the multi-order-parameters
interpenetration term (see Section~\ref{sec:meth:pf}), rather than explicitly (and arbitrarily) seeding $\beta$
nuclei at defined spatiotemporal locations. During the $\alpha' \to \alpha+\beta$ transformation, V diffuses
from $\alpha'$ martensite to the $\beta$ phase. Martensite grains transform into
$\alpha$ solely by the variation of their chemical composition. 

\subsubsection{Phase-field model}
\label{sec:meth:pf}
We use a standard formulation of the system free energy \cite{bkm:RefShi2015-77_1}:

\begin{align}
  F=\dfrac{1}{V_{m}} \int_{V} \bigg{[} f_c + \dfrac{\kappa_c}{2} |\nabla c|^{2} +\sum_{i=1}^{p} \bigg{(}\dfrac{\kappa_{\eta}}{2} |\nabla \eta_{i}|^{2} \bigg{)} \bigg{]} dV
  \label{eq:free_energy}
\end{align}
where $c$ is the V concentration, $f_c$ is the chemical free energy density, $\eta_{i}$ are the phase fields, $\kappa_{\eta}=3\sqrt{2}\sigma\delta V_{m}$ is the
gradient energy coefficient for $\eta_{i}$ \cite{bkm:RefLoginova2003-88}, with $\sigma$ the interface excess free energy,
$\delta$ the diffuse interface thickness, and $V_{m}$ is the molar volume of the alloy, $\kappa_c$ is the gradient energy coefficient for $c$,
and $V$ is the volume of the simulation domain.

This formulation does not include the effect of elastic strain energies, which may lead to anisotropic interfacial behaviour.
However, even though this may lead to morphological differences, since the kinetics of martensite decomposition in this alloy is limited by the diffusion of V 
\cite{bkm:RefHaubrich2019-22,bkm:RefTan2016-31,bkm:RefBarrioberoVila2017-34,bkm:RefAlBermani2010-63,bkm:RefSallicaLeva2016-64,bkm:RefKazantseva2018-65}, we assume that this does not have a significant effect on the transformation kinetics per se.
Also for the sake of simplicity, and since they are virtually identical in the model, both equilibrium ($\alpha$) and
martensitic ($\alpha'$) phases are simply referred to as phase $\alpha$ in the description of the model
below.

The chemical free energy density is \cite{bkm:RefZhu2004-89}:

\begin{align}
  f_c=f^{\alpha}h+f^{\beta}(1-h)+wg
\end{align}
where $f^{\alpha}$ and $f^{\beta}$ are the chemical free energy densities of $\alpha$ and $\beta$ phases, respectively, 

\begin{align}
  h=\sum_{i=1}^{p}\eta_{i}^{3}(6\eta_{i}^{2}-15\eta_{i}+10)
\end{align}
is an interpolation function, $w=6\sigma V_{m}/(\sqrt{2}\delta)$ is a parameter that controls the height of
the double well barrier \cite{bkm:RefLoginova2003-88}, and the function

\begin{align}
  g=\sum_{i=1}^{p}\bigg{[}\eta_{i}^{2}(1-\eta_{i})^{2}\bigg{]}+\psi \sum_{i=1}^{p}\bigg{[}\sum_{j \neq i}^{p}(\eta_{i}^{2}\eta_{j}^{2})\bigg{]}
\label{eq:g}
\end{align}
combines a standard double-well potential and a second term preventing the interpenetration of $\alpha$ grains of different
orientations, with $\psi=\xi \delta \sqrt{w/(2\kappa_{\eta})}$, where the coefficient $\xi$ parametrises the
free energy penalty associated with grain interpenetration (see, e.g. \cite{bkm:RefOforiOpoku2010-87}).

The chemical free energy densities of phases $\alpha$ and $\beta$ as a function of temperature and chemical
composition were computed by considering them as a regular solution with:

\begin{align}
  f^{\varphi}=&c_{Ti}\mu_{Ti}^{\varphi}+c\mu_{V}^{\varphi}+c_{Al}\mu_{Al}^{\varphi}+RT[c_{Ti}\ln{(c_{Ti})}+c\ln{(c)}+c_{Al}\ln{(c_{Al})}] \nonumber \\
  &+c_{Ti}c\sum_{k=0}^{n}[L_{TiV_{k}}^{\varphi}(c_{Ti}-c)^{k}]+c_{Al}c_{Ti}\sum_{k=0}^{n}[L_{AlTi_{k}}^{\varphi}(c_{Al}-c_{Ti})^{k}] \nonumber \\
  &+c_{Al}c\sum_{k=0}^{n}[L_{AlV_{k}}^{\varphi}(c_{Al}-c)^{k}]
\end{align}
for both phases $\varphi \in\{\alpha ,\beta \}$, where $\mu_{Ti}^{\varphi}$, $\mu_{V}^{\varphi}$, and $\mu_{Al}^{\varphi}$ are the
chemical potentials of Ti, V, and Al, respectively, $L_{TiV_{k}}^{\varphi}$, $L_{AlTi_{k}}^{\varphi}$, and $L_{AlV_{k}}^{\varphi}$
are Redlich-Kister coefficients, $R$ is the perfect gas constant, and $T$ is the temperature in Kelvin. 
Note that, given the chosen pseudo-binary description of the alloy (Fig.~\ref{bkm:FigPseudoBinary}), $c_{Al}$ is fixed, while $c_{Ti}$ is constrained to be equal to $1-c_{Al}-c$.
Chemical free energy densities for $\alpha$ and $\alpha'$ were computed with the same
equation but different local V concentrations. In the mixture term, the chemical potentials of the alloying elements, for the
different phases, were benchmarked against ThermoCalc results (using database TCNI8). The excess
term was computed by using standard Redlich-Kister polynomials with coefficients extracted from Ref.~\cite{bkm:RefAnsara1998-90} (see Appendix~\ref{appendix:chemfe}).
For the $\beta$-phase free energy density, we re-assessed the value of $L_{\mathit{AlTi}_0}^{\beta}=-118500+33.5\,T$, which corresponds to an increase by almost 5\% from the original value suggested by Ansara \textit{et al.} \cite{bkm:RefAnsara1998-90}. 
The modification of $L_{\mathit{AlTi}_0}^{\beta}$ was motivated by the resulting improved match between predicted $\beta$ volume fraction, also compared with ThermoCalc computation accounting for Al addition in the Ti-V binary system. 

Given the total free energy of the system, the evolution of V concentration and phase fields were computed using standard
Cahn-Hilliard and Allen-Cahn equations, respectively:

\begin{align}
  \dfrac{1}{V_{m}^{2}}\dfrac{\partial c}{\partial t} &= \nabla \cdot \bigg{[}\dfrac{M}{V_{m}}\nabla\bigg{(}\dfrac{\partial f_c}{\partial c} -\kappa_c\nabla^{2}c \bigg{)}\bigg{]}
  \label{eq:cahn-hilliard}
  \\
  \dfrac{\partial \eta_{i}}{\partial t} &= -\dfrac{L}{V_{m}}\bigg{(}\dfrac{\partial f_c}{\partial \eta_{i}}-\kappa_{\eta}\nabla^{2}\eta_{i}\bigg{)}
  \label{eq:allen-cahn}
\end{align}

\noindent where $M$ and $L$ correspond to solute and interface mobilities (see Section~\ref{sec:parameters}).

\subsubsection{Numerical resolution}
\label{sfftbr}

The simulation domain is spatially discretised in 2D and represents a  point of the AM  material, sufficiently small with respect to the macroscale to allow separation of scales and sufficiently large to contain a representative ensemble of grains. Even though the actual microstructures are not periodic, periodic boundary conditions (BCs) produce a faster convergence with the size of the domain to the actual response, and therefore we apply periodic BCs on both directions. 
Thus, the system of partial differential equations \eqref{eq:cahn-hilliard}-\eqref{eq:allen-cahn} is solved
in a monolithic way using Fourier spectral method with a first-order finite difference scheme for time discretisation.

Due to the complexity and stringent stability requirements of the fourth-order partial differential equation, the use of a non-implicit approach for
Cahn-Hilliard equation resolution is not robust and eventually diverges.
Therefore, the Cahn-Hilliard equation is solved by using a fully-implicit iterative method, described in Appendix~\ref{appendix:CH}. 
The Allen-Cahn equation, on the other hand, can be solved by using a semi-implicit non-iterative algorithm, as described in Appendix~\ref{appendix:AC}, as
the second-order differential equation is inherently more stable. The resulting discretised system of equations form a system of two algebraical equations,
as detailed in Appendix~\ref{appendix:resol}. The resolution is accelerated by the use of a preconditioned conjugate gradient method (PCG) \cite{barrett_1994},
following a procedure presented in \cite{LUCARINI2019103131}, with a variable time step (see Appendix~\ref{appendix:resol}).

A rectangular domain of size $l_{x} \times l_{y}$ is discretised into a regular grid containing $p_{x}$ and $p_{y}$ points
in $x$ and $y$ directions, respectively. In order to reduce the aliasing effect in the presence of non-smooth functions,
the discrete Fourier frequency vector and the square of the frequency gradient are computed by replacing the definition
of the derivative in the Fourier space with a fourth-order finite difference computed through the use of Fourier transform. 
This procedure results in the redefinition of the frequencies \cite{bkm:RefBoccardo2023-91}:

\begin{align}
  \boldsymbol{\xi}=&\bigg{[}\dfrac{8 \sin(2 \pi n_{x} \Delta x/l_{x})-\sin(4 \pi n_{x} \Delta x /l_{x})}{6 \Delta x}, \nonumber \\
    & \ \ \dfrac{8 \sin(2 \pi n_{y} \Delta y/l_{y})-\sin(4 \pi n_{y} \Delta y/l_{y})}{6 \Delta y}\bigg{]}
  \label{eq:grad_fft-fdO4}
  \\
  \|\boldsymbol{\xi}\|^{2} =& \bigg{[}\dfrac{\cos(4 \pi n_{x} \Delta x/l_{x})-16\cos(2 \pi n_{x} \Delta x/l_{x})+15}{6 \Delta x^{2}} + \nonumber \\
    & \ \ \dfrac{\cos(4 \pi n_{y} \Delta y/l_{y})-16\cos(2 \pi n_{y} \Delta y/l_{y})+15}{6 \Delta y^{2}}\bigg{]}
  \label{eq:Gamma_fft-fdO4}
\end{align}
where $n_{x}$ and $n_{y}$ are two-dimensional mesh grid matrices generated with two vectors of the form $[0,...,(p_{x}/2),-(p_{x}/2-1),...,-1]$ and
$[0,...,(p_{y}/2),-(p_{y}/2-1),...,-1]$, and $\Delta x$ and $\Delta y$ are the distance between two consecutive points in the $x$ and $y$ directions, respectively.
While, for the sake of generality, the equations above are presented for any $\Delta x\neq\Delta y$, here all application are with  $\Delta x=\Delta y$.

The computational scheme, summarised in Algorithms~\ref{alg:fft-based_solution} and \ref{alg:c_solution} of Appendix~\ref{appendix:algo}, 
is implemented using the Python programming language. For each time step, the computation of discrete Fourier
transform ($\mathcal{F}$) and discrete inverse Fourier transform ($\mathcal{F}^{-1}$) is performed on the GPU device
using Scikit-CUDA \cite{sk_cuda_2021}. To solve a system of algebraical equations on the GPU device, CUDA kernels are
programmed through PyCUDA \cite{kloeckner_pycuda_2012}, where the arrays representing the field values in the domain
are defined in double precision. 

The simulations were performed using a single GPU on a computer with the following
hardware features: Intel Xeon Gold 6130 microprocessor, 187~GB RAM, GeForce RTX 2080Ti GPU (4352 Cuda cores and 11~GB RAM), and
software features: CentOS Linux 7.6.1810, Python 3.8, PyCUDA 2021.1, Scikit-CUDA 0.5.3, and CUDA 10.1 (Toolkit 10.1.243).
The GPU block size was set to $16\times 16$, which was found to result in near-optimal performance.

\subsubsection{Alloy and model parameters}
\label{sec:parameters}

The molar volume of the alloy was assumed to be independent of temperature and equal to that of pure Ti, i.e. 
$V_m=1.064\times 10^{-5}~$m$^3$/mol \cite{bkm:RefSingman1984-92}. The excess free energy of interfaces was
taken as $\sigma =0.1~$J/m$^2$ \cite{bkm:RefAhluwalia2020-80} and the interpenetration of different $\alpha$ grains was
prevented using a coefficient $\xi =1000$. 
The chemical mobility is assumed isotropic, but phase/location-dependent, hence 
computed as $M=(1/V_{m})[(1-c)^{2}cM_{V}+c^{2}(c_{Al}M_{Al}+c_{Ti}M_{Ti})]$, where the individual atomic
mobilities $M_{k}$ (with $k{\in}\{V,Al,Ti\}$) depend upon the phase as
$M_{k}=M_{k}^{\alpha}+M_{k}^{\beta}-M_{k}^{\alpha (1-\sum_{i=1}^{p}\eta_{i})}M_{k}^{\beta(\sum_{i=1}^{p}\eta_{i})}$. The atomic
mobilities $M_{k}^{\alpha}$ and $M_{k}^{\beta}$ and their dependence upon temperature and V concentration, were modelled using
Redlich-Kister polynomials with coefficients extracted from Ref.~\cite{bkm:RefGierlotka2019-93} (see Appendix~\ref{appendix:atomicm}). Like in the computation of the chemical free
energy, it was assumed that the atomic mobilities of $\alpha$ and $\alpha'$ could be computed with the same
equation. The interfacial mobility, assumed isotropic and independent of grain orientations, was computed as
$L=0.00125\exp\{-19500/T\}~$m$^3$/(Js) \cite{bkm:RefAhluwalia2020-80}, which was adjusted to
allow the simulated $\beta$ volume fraction to reach the steady stable condition at the different plateau temperatures,
in agreement with experiments and literature \cite{bkm:RefMurgau2012-74}. The phase field diffuse interface width was taken as
$\delta =10\Delta x$.

\subsubsection{Model validations}
\label{sec:method:valid}

Before comparing the PF results with our experiments, we validated the predictions in terms of thermodynamics (equilibrium) and
transformation kinetics of the model.
All validation simulations considered the initial microstructure of region~1 (see Section~\ref{sec:methods:simu}, Figure~\ref{bkm:FigICs}). 

To validate the prediction of the equilibrium state, we simulated the isothermal annealing at several temperatures between 650$^\circ$C and 980$^\circ$C.
The annealing time was large enough to reach equilibrium (i.e. full $\alpha'\rightarrow\alpha+\beta$ transformation when relevant). 
Then, we compared PF-predicted results (namely the volume fraction of $\beta$ phase at different temperatures) with theoretical equilibrium (lever rule), with our experiments, as well as with a polynomial fit to experimental data from the literature (however, for traditionally manufactured Ti-6Al-4V) \cite{bkm:RefMurgau2012-74}. 
The equilibrium concentrations of $\alpha$ and $\beta$ phases used in the lever rule calculation were obtained by the common tangent construction from $\alpha$ and $\beta$ free energy densities. 

In order to validate the transformation kinetics with the identified parameters, we compared the results of the model to the
martensite decomposition kinetics assessed experimentally by Gil Mur \textit{et al.} \cite{bkm:RefGilMur1996-73}. Their results correspond to a wrought Ti-6Al-V alloy, annealed at 1050$^{\circ}$C for 30 minutes and water quenched at room temperature in order to obtain a fully $\alpha'$ martensitic microstructure before the annealing experiment. 
The sample hardness, measured at different annealing times and temperatures, is used as a proxy for the fraction evolution in the assessment of Avrami exponents of the $(\alpha'\rightarrow\alpha+\beta)$ transformation at different temperatures. 
Using the hardness as a stand-in for the phase fraction is a relatively strong assumption, since it neglects possible additional phenomena, such as a potential recovery process. While acknowledging this limitation and hence only using it as an order-of-magnitude estimate, we use this data because it is, to the best of our knowledge, the only available regarding the kinetics of the martensite decomposition at 700$^{\circ}$C and 800$^{\circ}$C.
We discarded reported temperature of 400$^{\circ}$C and 600$^{\circ}$C to focus on cases with complete transformation by the end of annealing. 

Because the thermal history is not clearly specified in \cite{bkm:RefGilMur1996-73}, we consider a uniform
temperature distribution with two different thermal paths: a) isothermal at the given temperature and b) with a heating ramp from
room temperature up to the given temperatures. 
The nonlinear heating ramp was simulated with the thermal module of the finite element software
Abaqus \cite{smith_2009}, considering a cylindrical sample of 3~mm in length and 5~mm in diameter, with a density 4430~kg/m$^{3}$,
specific heat capacity 526.3~J/(kg$^{\circ}$C) and thermal conductivity 6.7~W/(m$^{\circ}$C) \cite{bkm:RefBoyer1994-20}.
The boundary condition imposed at the cylinder interface follows a Newton law, $q_{c}=-h_{c}(T-T_{env})$,
where $q_{c}$ is the normal heat flux, $h_{c}$ is the heat transfer coefficient at interface, $T$ is the temperature
at the sample interface, and $T_{env}$ is the environment/annealing temperature. The heat transfer coefficient
$h_{c}=70~$W/(m$^{2}$K) corresponds to a regular furnace heating \cite{boccardo_2017}. The temperature evolution
of the cylinder, initially at room temperature, was extracted from the centre of the cylinder (but the temperature is
almost uniform in all the sample) and imposed, as an input data of the PF model.

\subsubsection{Simulations}
\label{sec:methods:simu}

Ultimately, we aim to simulate the microstructural evolution of the \textit{in-situ} observed microstructures during the step-wise annealing process (Figure~\ref{bkm:FigTempVsTime}) and compare it with our \textit{in-situ} experiments. 
Hence, we use experimentally characterised microstructures as initial
conditions. In particular, two different regions of the fully martensitic as-built microstructure were separately used as the
initial material microstructure. Both selected regions correspond to a single parent
$\beta$ grain, each hosting several $\alpha'$ variants (orientations).

\begin{figure}[t!]
  \begin{center}
    \includegraphics[width=3.1in]{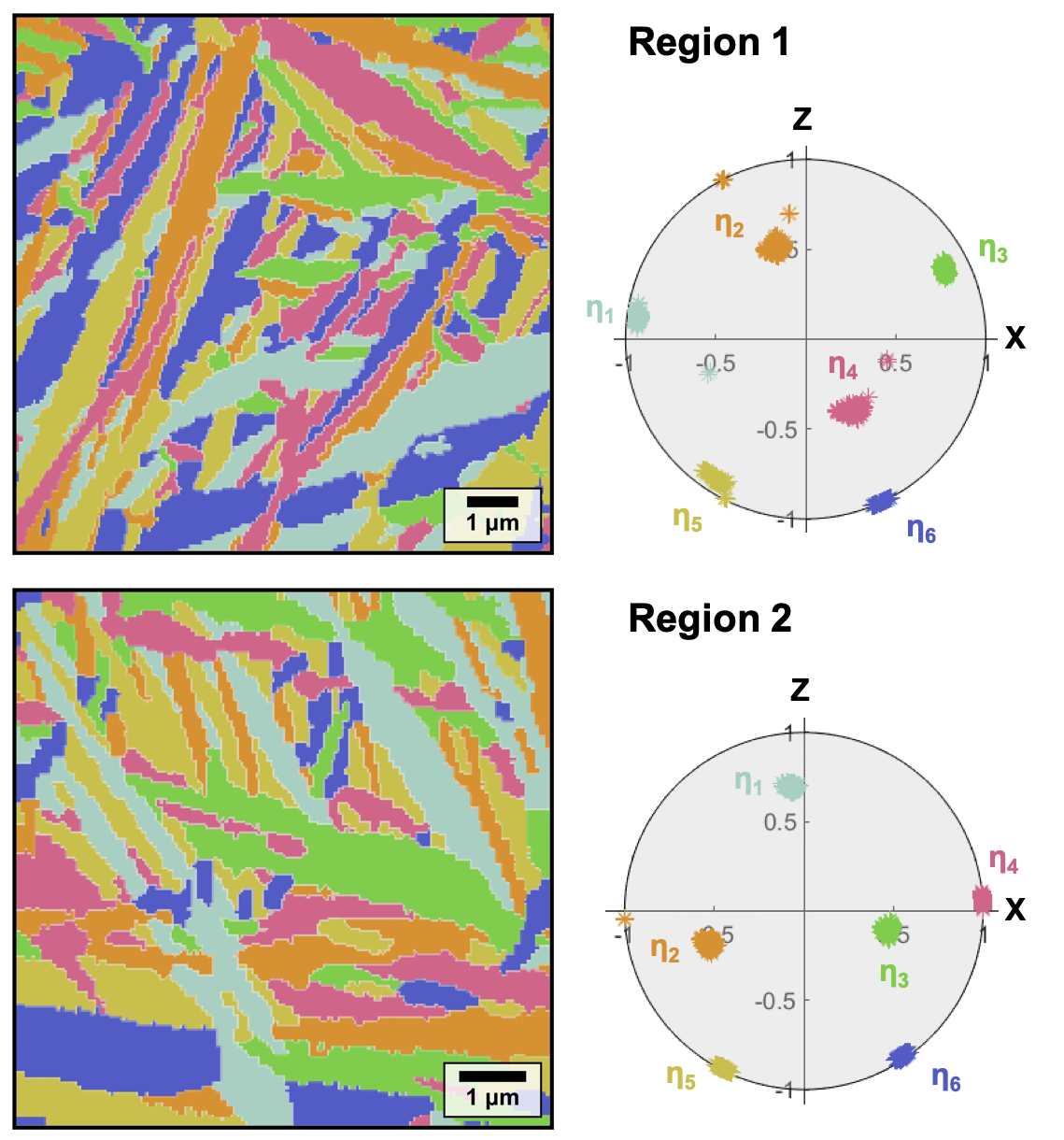}
    \caption{Initial conditions for the phase fields, based on EBSD maps in the as-printed material, showing region 1 and region 2. Colours represent the six different crystal orientations of $\alpha'$ considered in the simulation.
      }
    \label{bkm:FigICs}
  \end{center}
\end{figure}

Our model distinguishes regions of different orientations, but it does not attribute a specific given orientation to the grains. 
In other words, neighboring grains of different orientations result in grain boundaries (and subsequent grain boundary energies), but the  behavior of the grain and grain boundaries is isotropic.
We classified the different $\alpha'$ orientations into variants, for each prior-$\beta$ grain. To sort the different $\alpha'$
variants, we used experimental EBSD maps to identify groups of hcp grains with respect to their $c$-axis orientation, using the
following procedure, illustrated in Figure~\ref{bkm:FigICs}. For each pixel in the EBSD map, the tip of the unit vector oriented along the local crystalline $c$-axis and rooted at the origin of a Cartesian coordinated system $(x,y,z)$ was orthogonally projected onto the $(x,z)$ plane. 
The resulting clouds of points, visible on the right-hand-side of Figure~\ref{bkm:FigICs} for both investigated regions, exhibit clusters of orientations, corresponding to a subset of the 12 possible variants \cite{bkm:RefWielewski2012-26,bkm:RefSimonelli2014-27,bkm:RefHumbert1995-37,bkm:RefHumbert1996-38,bkm:RefGlavicic2003-39,bkm:RefGlavicic2003b-40}.

Here, we clearly found that 6 main clusters could be identified in each prior-$\beta$ grain. For that reason, we considered that accounting for 6 variants in each prior-$\beta$
grain was sufficient, assigning the same large cluster variant to the nearest small clusters. The resulting initial microstructures, composed of 6 $\alpha '$
lath orientations, are represented on the left-hand-side of Figure~\ref{bkm:FigICs} -- each colour corresponds to one of the identified orientations, and hence one of the phase fields  $\eta_{i}$.

At the start of the simulations, the V concentration is assumed uniform in the entire domain, and equal to the nominal V
concentration of the alloy. 
This is consistent with experimental observations showing that, under powder-bed fusion processing, only $\alpha'$ was formed at a chemical composition close to the nominal alloy concentration \cite{bkm:RefKazantseva2018-65}.
The experimental temperature-time profile (Figure~\ref{bkm:FigTempVsTime}) was used as input,
considering a spatially homogeneous but time-dependent temperature.

Both considered regions are spatially discretised with a structured grid of $256\times256$ points, resulting in grid spacings
$\Delta x=\Delta y$ of 41.0~nm for region 1 of size (10.5~\textmu m)$^{2}$, or 31.3~nm for region 2 of size (8.0~\textmu m)$^{2}$.

%
\section{Results}
%

\subsection{Experiments}
\label{sec:resu:exp}

\begin{figure}[t!]
  \begin{center}
    \includegraphics[width=5.7in]{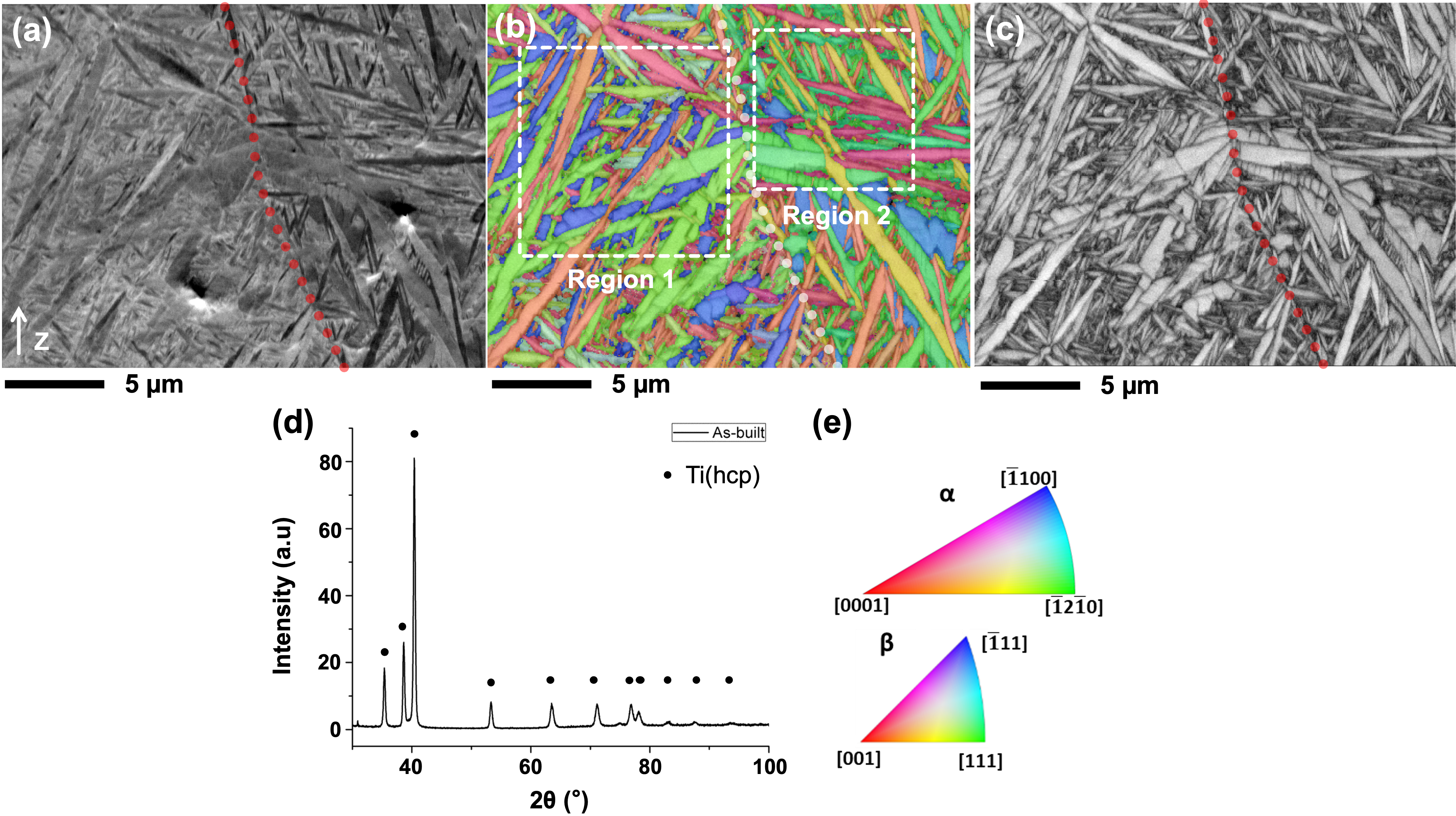}
    \caption{Electron micrographs of as-printed microstructure of L-PBF
      Ti-6Al-4V: (a) microstructure image captured by forescatter detectors; (b) measured $\alpha$ orientation map in Z-IPF overlaid on band contrast image; (c) measured $\beta$ orientation map in Z-IPF overlaid on band contrast image (here, highlighting the absence of $\beta$ phase); (d) XRD results of the crystal lattices; (e) colour scheme used to map the orientation data.
Semi-transparent dotted lines in (a)-(c) mark the location of the prior-$\beta$ grain boundary.
      }
    \label{bkm:FigExptsICs}
  \end{center}
\end{figure}

The microstructure in as-built L-PBF Ti-6Al-4V in the investigated area is shown in Figure~\ref{bkm:FigExptsICs}. As discussed in previous works \cite{zou2021microstructure}, LPBF leads to a fully acicular $\alpha'$ martensitic microstructure, 
(Figure~\ref{bkm:FigExptsICs}(a) and (b)), with no detectable retained $\beta$ (Figure~\ref{bkm:FigExptsICs}(c)). 
The XRD results in Figure~\ref{bkm:FigExptsICs}(d) further confirm the absence
of the $\beta$ phase in the microstructure. The prior-$\beta$ structure can also be identified from
forescatter electron micrographs. 
A prior-$\beta$ grain boundary is indicated by dotted lines in Figs~\ref{bkm:FigExptsICs}(a)-(c).
Multiple $\alpha'$ grains of distinct orientations were visible in the prior-$\beta$ microstructure, and reveal a
weak crystallographic texture. 

\begin{figure}[t!]
  \begin{center}
    \includegraphics[width=5.45in]{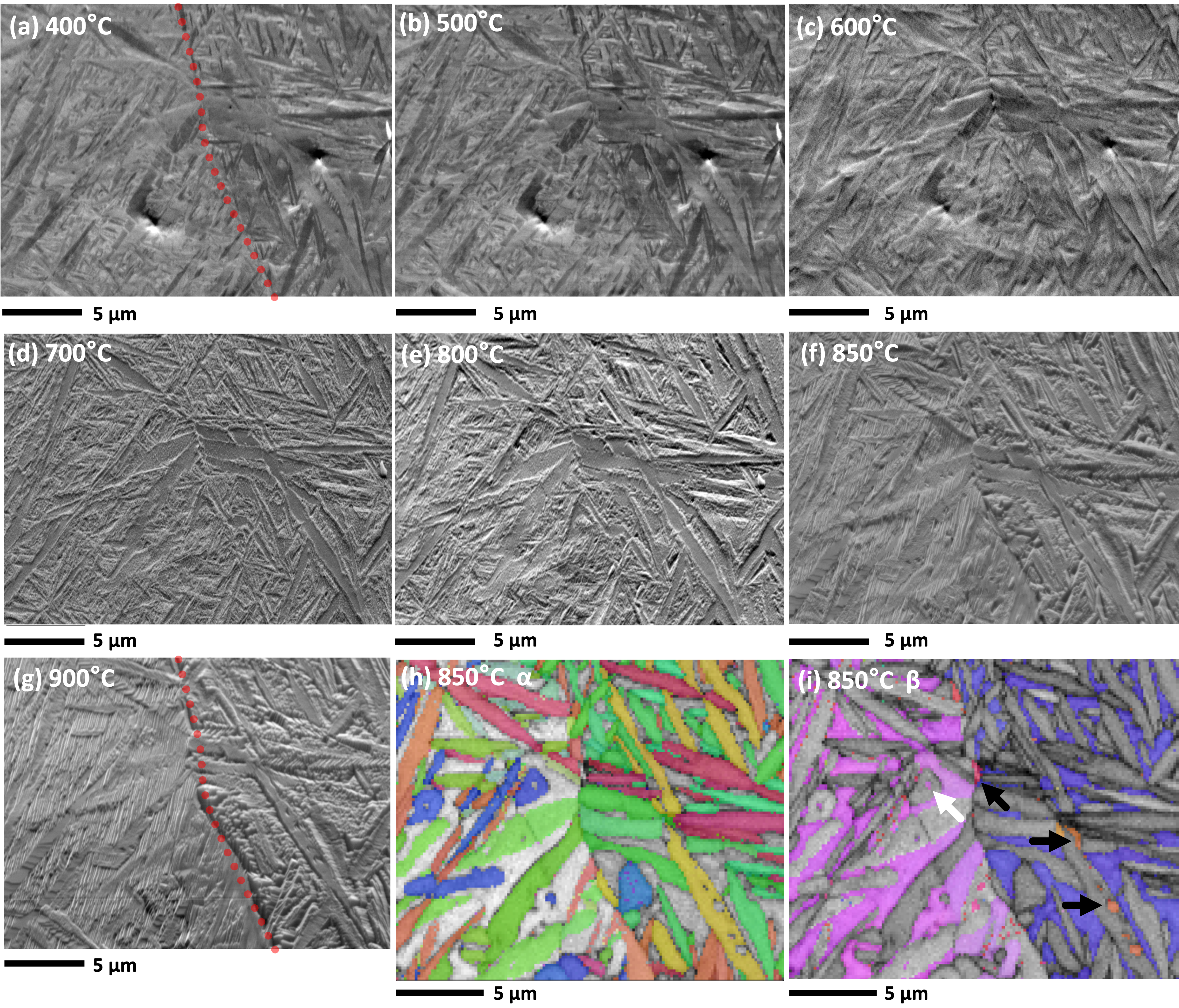}
    \caption{SEM images of microstructure evolution of L-PBF Ti-6Al-4V under stepwise heating (Figure~\ref{bkm:FigTempVsTime}). 
      Forescatter micrographs at (a) 400$^{\circ}$C, (b) 500$^{\circ}$C, and (c) 600$^{\circ}$C. 
      Secondary electrons micrographs at (d) 700$^{\circ}$C, (e) 800$^{\circ}$C, (f) 850$^{\circ}$C, and (g) 900$^{\circ}$C.
      Measured $\alpha$ (h) and $\beta$ (i) orientation map in Z-IPF (overlaid
      with band contrast image and using the colour scheme of Fig.~\ref{bkm:FigExptsICs}e), showing the crystallographic texture at 850$^{\circ}$C, with a white arrow indicating the
      $\beta$ phase growth from the interior of the $\alpha'$ phase, and black arrows indicating the recrystallised high
      angle boundary (HAB) $\beta$ phase.
      The red dotted line in (a) and (g) marks the location of the initial prior-$\beta$ GB.
      }
    \label{bkm:FigExptsVsTime}
  \end{center}
\end{figure}

Figure~\ref{bkm:FigExptsVsTime} shows \textit{in-situ} micrographs captured during the stepwise thermal treatment (Figure~\ref{bkm:FigTempVsTime}).
Up to 600$^{\circ}$C, Figure~\ref{bkm:FigExptsVsTime}(a)-(c) display forescatter electron images.
At 700$^{\circ}$C and above  (Figure~\ref{bkm:FigExptsVsTime}(d)-(g)), the forescatter detectors were completely saturated, and the microstructure image was then captured by secondary electron detectors. 
The last two panels highlight the crystallographic texture at 850$^{\circ}$C, via the $\alpha$ (h) and $\beta$ (i) orientation map in Z-IPF overlaid with band contrast image.
The two distinct prior-$\beta$ grains also appear clearly in Figure~\ref{bkm:FigExptsVsTime}(f) to (i), corresponding to high temperatures, yet not high enough to lead to substantial evolution of the prior-$\beta$ grain structure within the considered time.
Moreover, while we only present here the EBSD maps in the as-built state (Fig.~\ref{bkm:FigExptsICs}) and at 850$^\circ$C (Fig.~\ref{bkm:FigExptsVsTime}) after the stepwise heat treatment, further experimental characterization can be found in previous publications (e.g. \cite{bkm:RefZou2020-51}).

\subsection{Model \& parameters validations}
\label{sec:resu:exp}

\begin{figure}[h!]
  \begin{center}
    \includegraphics[width=3.2in]{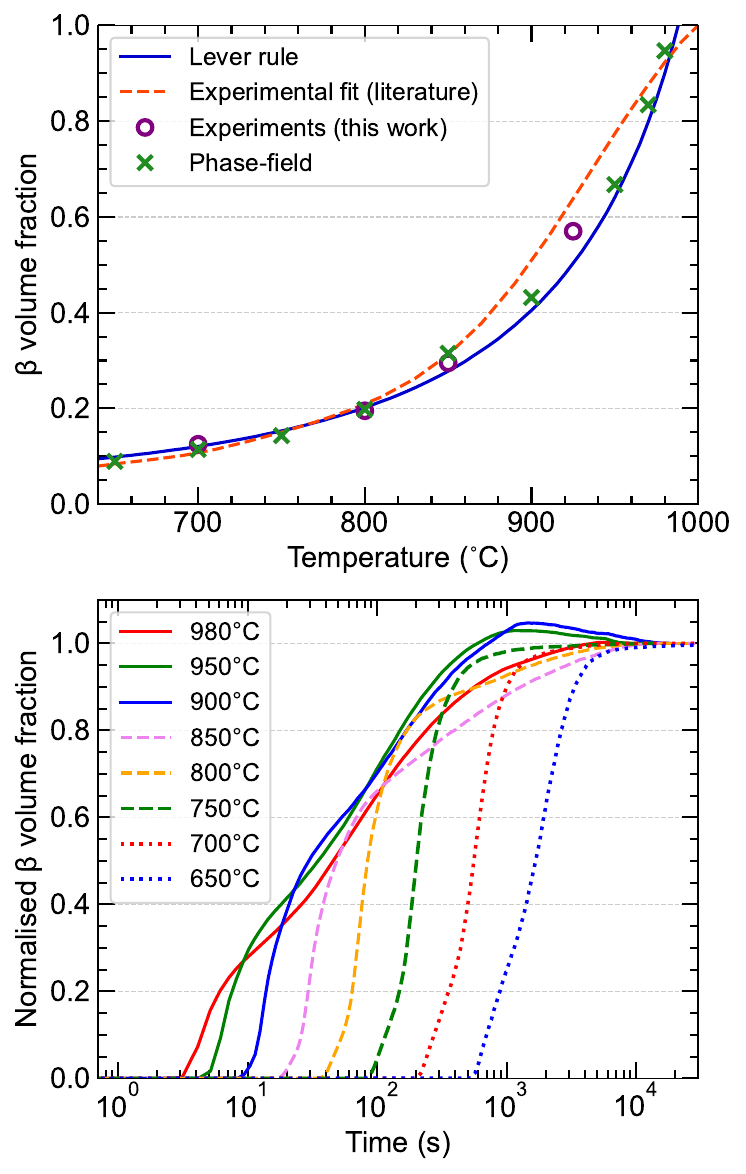}
    \caption{(Top) Equilibrium $\beta$ volume fraction of Ti-6Al-4V at high temperature, comparing lever rule,
      experimental fit for traditionally manufactured Ti-6Al-4V \cite{bkm:RefMurgau2012-74},
      phase-field results, and our experimental results for L-PBF Ti-6Al-4V. 
      (Bottom) Kinetics of $\beta$ transformation
      computed with the phase-field model and normalised with respect to the final $\beta$ fraction.
      }
    \label{bkm:FigLEVERvsPFvsExpts}
  \end{center}
\end{figure}

Figure~\ref{bkm:FigLEVERvsPFvsExpts} (top) shows the simulated evolution of the $\beta$ volume fraction during isothermal annealing of region~1 at different temperatures (green crosses) compared with our experimental data (purple circles), the lever rule (solid blue line), and the experimental fit for traditionally manufactured Ti-6Al-4V (red dashed line) \cite{bkm:RefMurgau2012-74}. 
As expected for a nominal Ti-6Al-4V across this temperature
range \cite{bkm:RefKelly2004-102,bkm:RefIdhil2016-103}, the $\beta$ volume fraction increases with temperature. 
Figure~\ref{bkm:FigLEVERvsPFvsExpts} (bottom) also includes the evolution of $\beta$ volume fraction, computed with the PF model, normalised with respect to the final $\beta$ volume fraction at each temperature. 

\begin{figure}[!t]
  \begin{center}
    \includegraphics[width=3.1in]{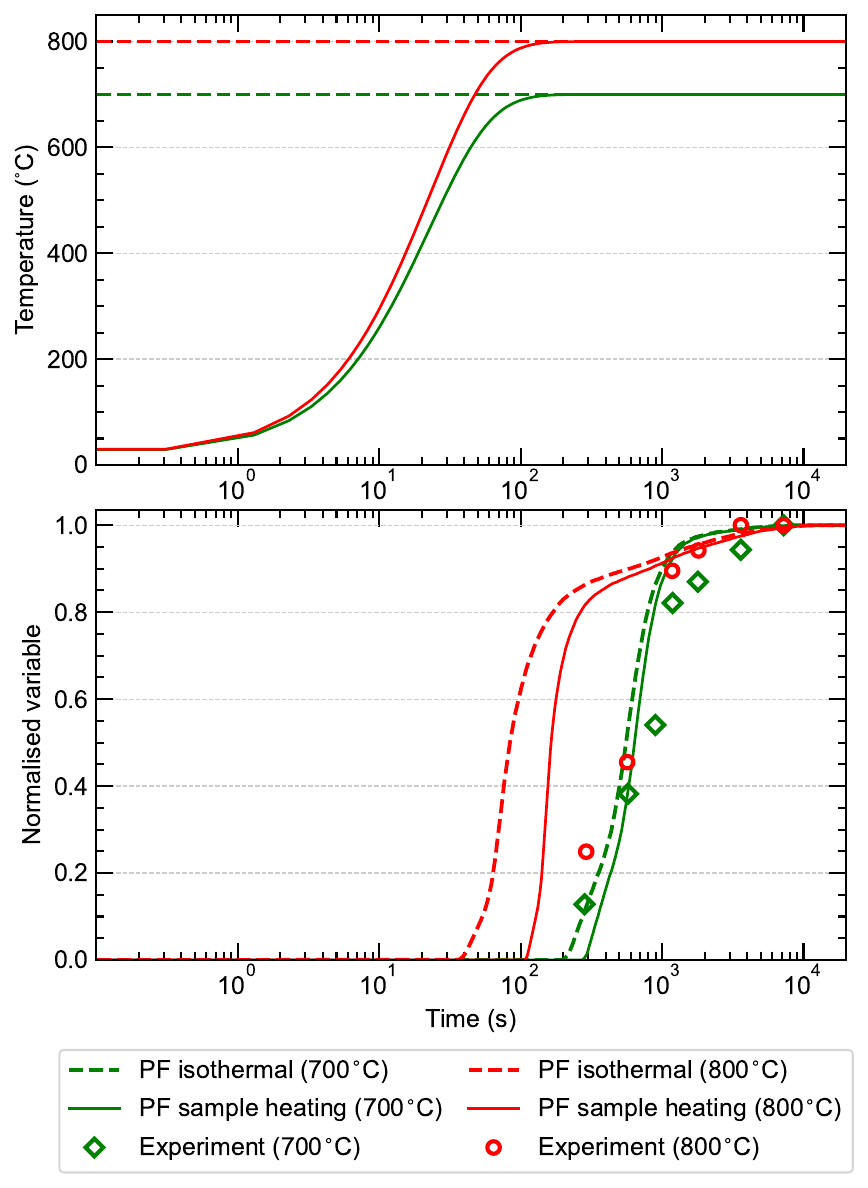}
    \caption{Temperature evolution (top) and kinetics of $\alpha'\rightarrow\alpha+\beta$ transformation computed with the phase-field model for region~1 (bottom), compared with experimental data \cite{bkm:RefGilMur1996-73} ($y-$axis: normalised $\beta$ fraction from PF simulations; normalised hardness from experiments).
    }
    \label{bkm:gilmur}
  \end{center}
\end{figure}

Figure~\ref{bkm:gilmur} presents the evolution of the temperature (top) and the transformation kinetics (bottom) at 700$^\circ$C and 800$^\circ$C for the two considered thermal paths, namely fully isothermal or accounting for the heating up of the sample (see Section~\ref{sec:method:valid}).
The transformed $\beta$ fraction is compared to the (hardness-based) experimental data (symbols) from the literature \cite{bkm:RefGilMur1996-73}. 
In order to compare the different data sets (namely, hardness from experiments \cite{bkm:RefGilMur1996-73} and $\beta$ fraction from our PF simulations), at each temperature we
compute the normalised $\beta$ fraction $\overline{f_{\beta}}=f_{\beta}/f_{\beta}^{end}$ and normalised hardness $\overline{H}=(H-H^{min})/(H^{max}-H^{min})$, where $f_{\beta}^{end}$ is
the $\beta$ fraction at the end of the PF simulation, while $H^{max}$ and $H^{min}$ are the maximum and minimum measured hardness, respectively.

%

\subsection{Phase-field simulations of microstructure evolution}
\label{sec:resu:exp}

\begin{figure}[t!]
  \begin{center}
    \includegraphics[width=3.5in]{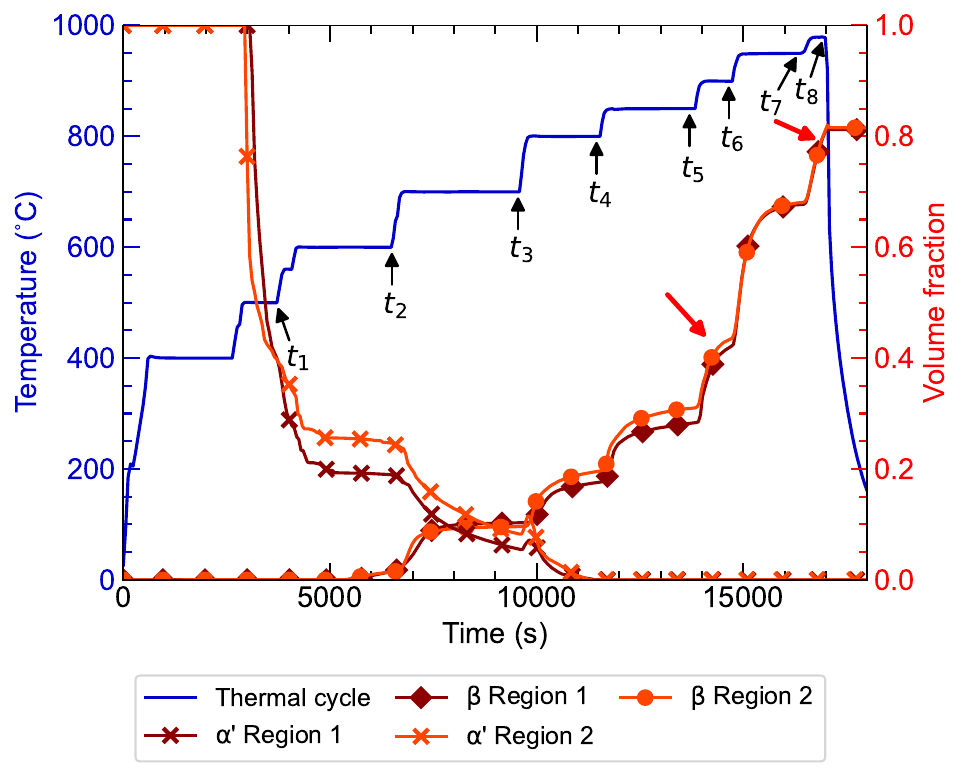}
    \caption{Temperature (left) and $\alpha'$ and $\beta$ volume fraction (right) evolution
      during the heat treatment.}
    \label{bkm:FigFracVsTime_1}
  \end{center}
\end{figure}

\begin{figure}[t!]
  \begin{center}
    \includegraphics[width=5.7in]{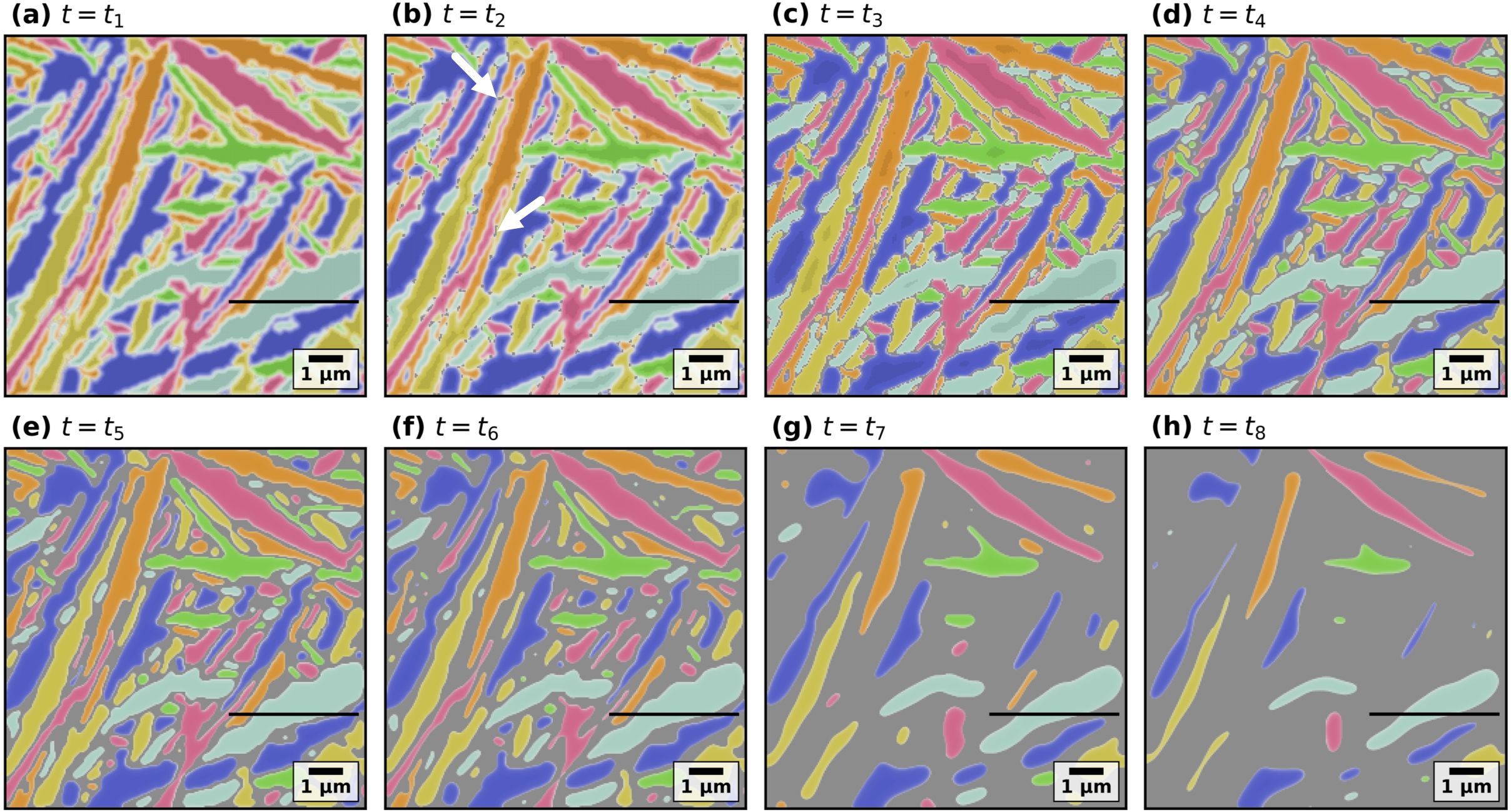}
    \caption{Simulation results of material microstructure in region 1
      at the end of the isothermal plateaus at (a) 500$^{\circ}$C, (b) 600$^{\circ}$C,
      (c) 700$^{\circ}$C, (d) 800$^{\circ}$C, (e) 850$^{\circ}$C, (f) 900$^{\circ}$C, (g) 950$^{\circ}$C, and (h) 980$^{\circ}$C;
      colour maps show different phases and grains ($\beta$: grey, $\alpha$: different colours
      for different variants, and $\alpha'$: different shaded colours for different variants). 
      The horizontal black line marks the location of the compositional line-scan of Figure~\ref{bkm:V_concentration_line}.
      White arrows in (b) show some $\beta$ nucleation sites along $\alpha'$ martensite lath boundaries.}
    \label{bkm:FigPFregion1}
  \end{center}
\end{figure}

\begin{figure}[t!]
  \begin{center}
    \includegraphics[width=5.7in]{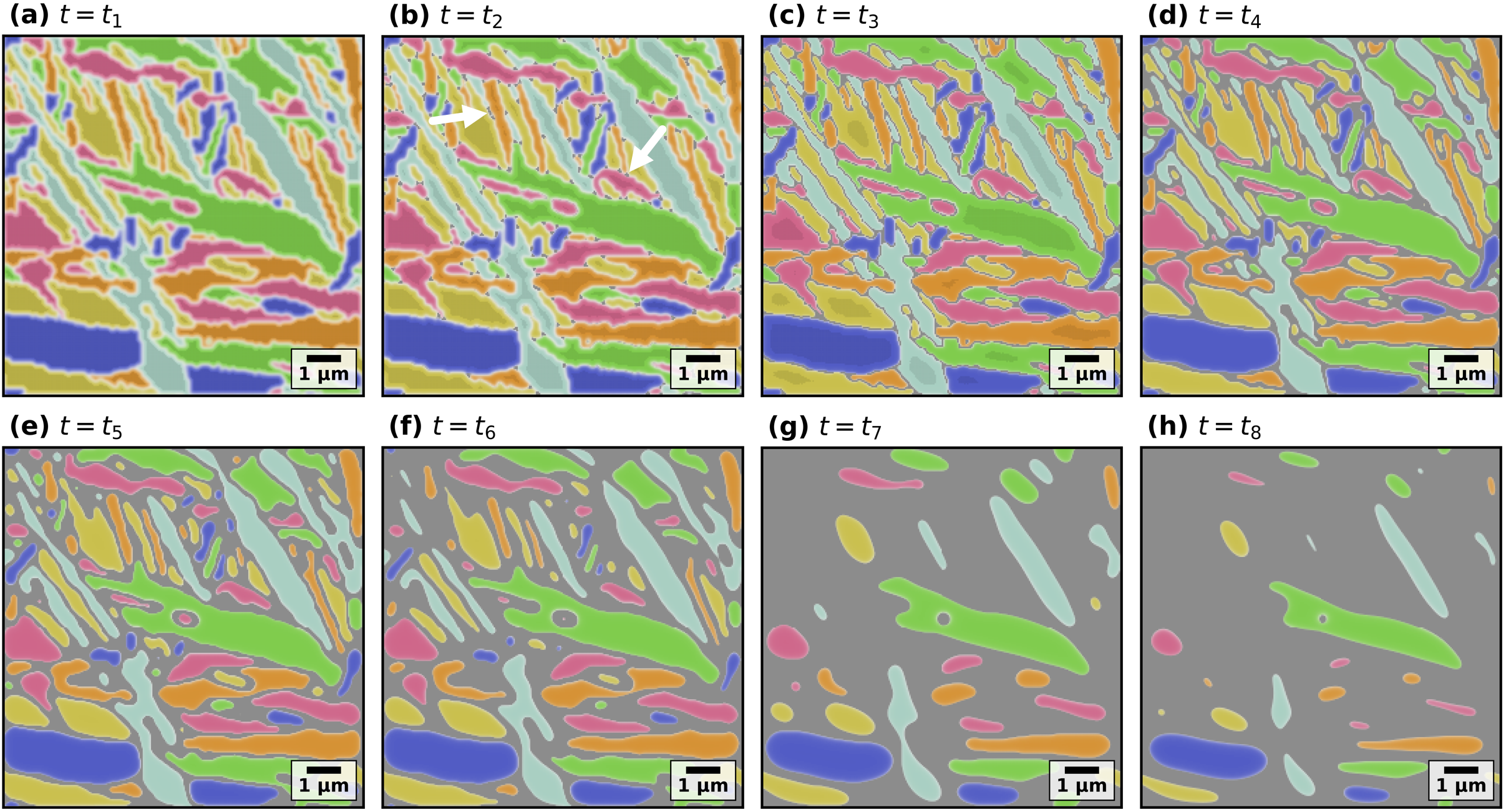}
    \caption{Simulation results of material microstructure in region 2
      at the end of the isothermal plateaus at (a) 500$^{\circ}$C, (b) 600$^{\circ}$C,
      (c) 700$^{\circ}$C, (d) 800$^{\circ}$C, (e) 850$^{\circ}$C, (f) 900$^{\circ}$C, (g) 950$^{\circ}$C, and (h) 980$^{\circ}$C;
      colour maps show different phases and grains ($\beta$: grey, $\alpha$: different colours
      for different variants, and $\alpha'$: different shaded colours for different variants).
      White arrows showing some $\beta$ nucleation sites along $\alpha'$ martensite lath boundaries.}
    \label{bkm:FigPFregion2}
  \end{center}
\end{figure}

\begin{figure}[t!]
  \begin{center}
    \includegraphics[width=3.7in]{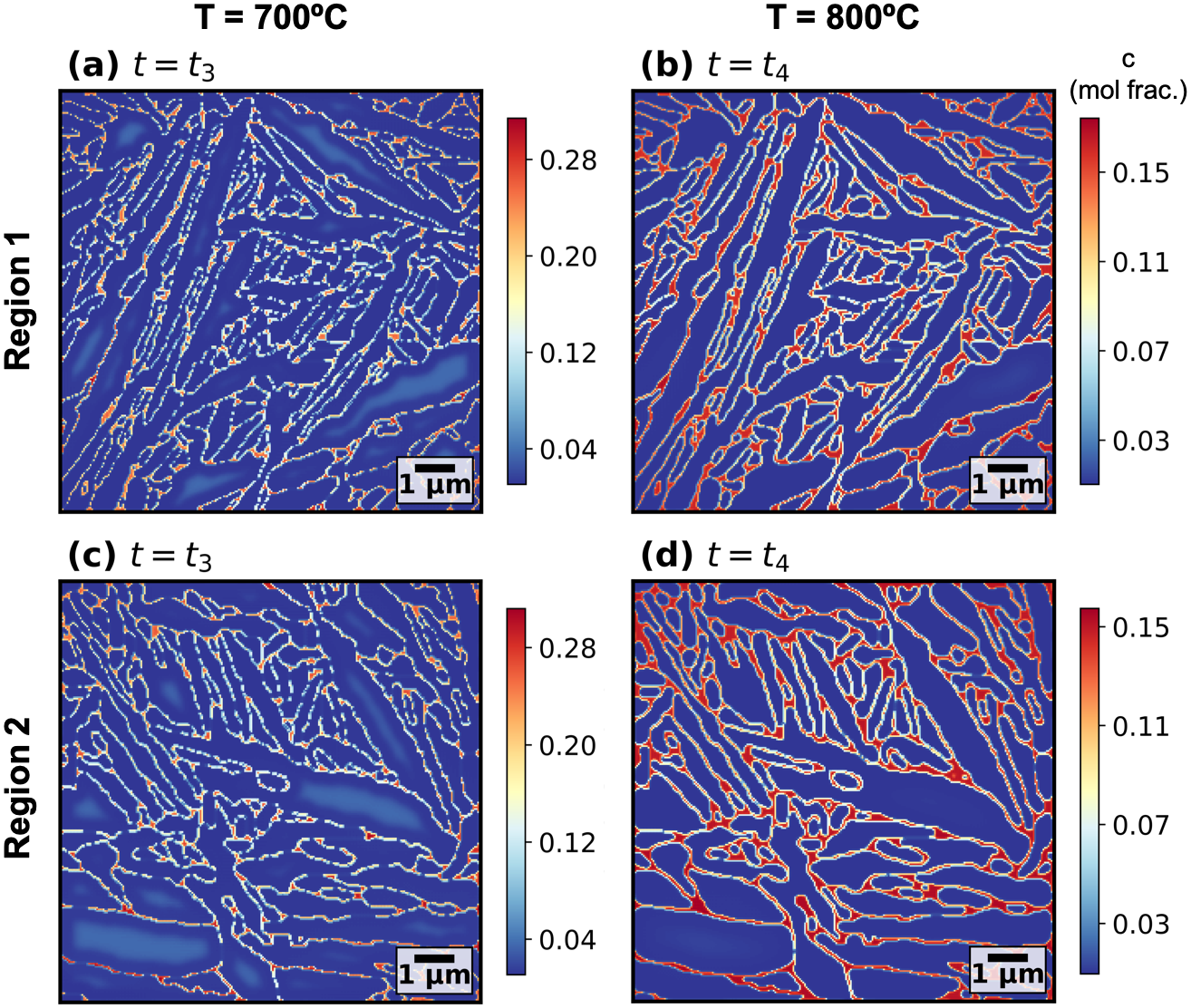}
    \caption{Simulation results of V concentration, in mole fraction, for regions 1 (a,b) and 2 (c,d)
      at the end of the isothermal plateaus at 700$^{\circ}$C (a,c) and 800$^{\circ}$C (b,d).}
    \label{bkm:FigPFregions1-2_conc}
  \end{center}
\end{figure}

\begin{figure}[h!]
  \begin{center}
    \includegraphics[width=4.8in]{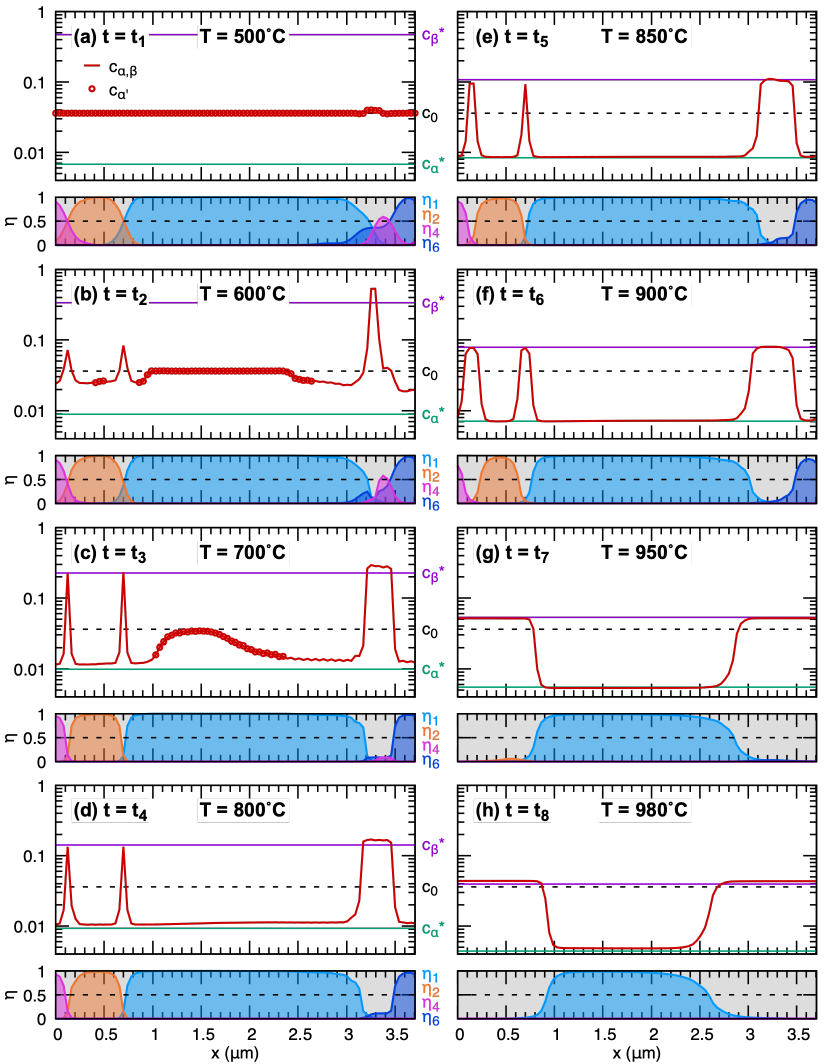}
    \caption{Compositional scans along the black horizontal line of region 1 in Figure~\ref{bkm:FigPFregion1} at different times/temperatures. 
    Each panel (a-h) shows (top) the V concentration, in mole fraction (red line, with symbols for $\alpha'$ regions) compared to nominal, $c_0$, and equilibrium, $c_\alpha$ and $c_\beta$, concentrations, as well as (bottom) order parameters, $\eta_i$, along the same line.}
    \label{bkm:V_concentration_line}
  \end{center}
\end{figure}

\begin{figure}[h!]
  \begin{center}
    \includegraphics[width=3.3in]{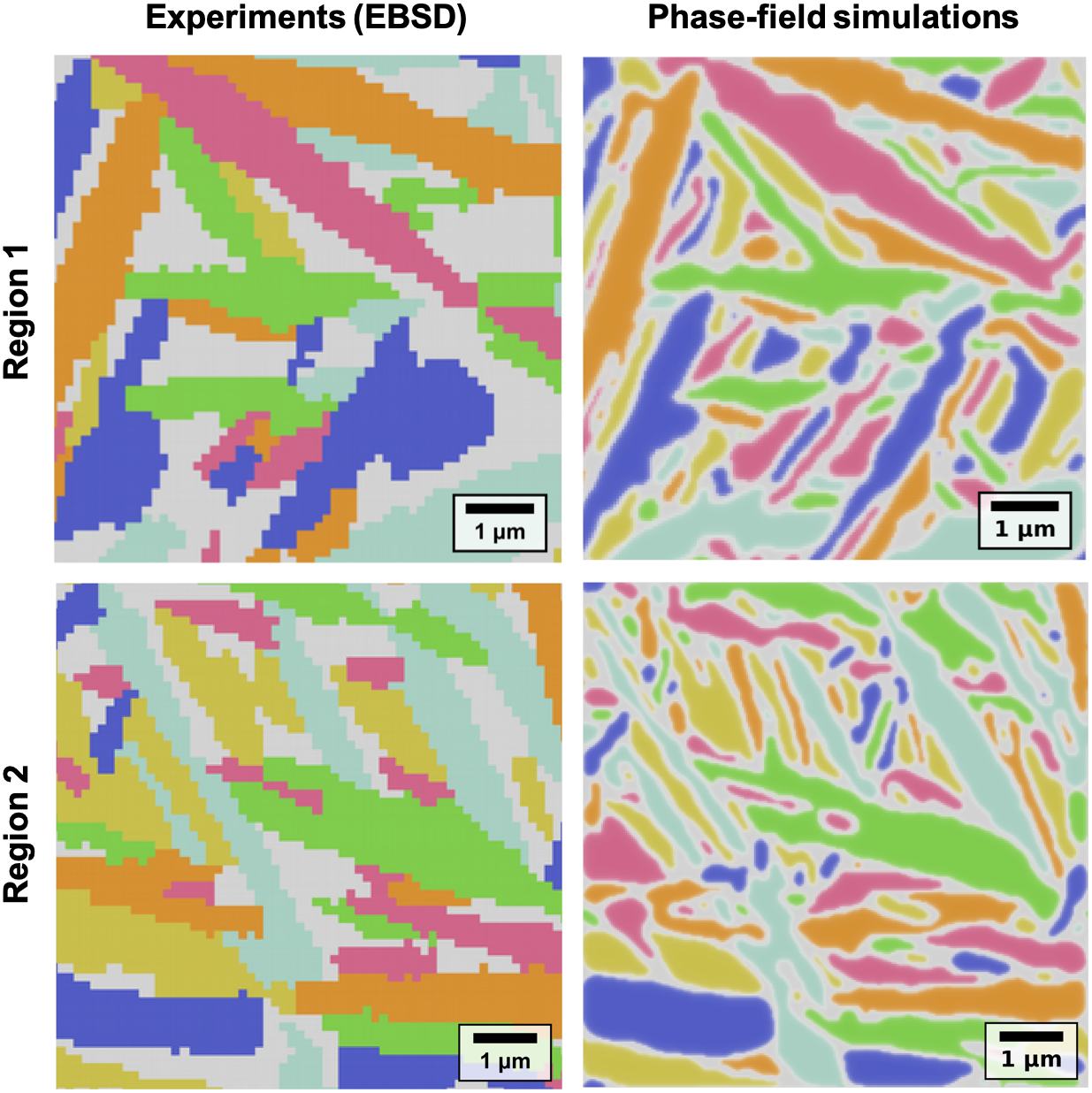}
    \caption{Computational simulation results compared to experimental microstructures in region 1 and region 2, after 300\,s of isothermal annealing at 850$^{\circ}$C.}
    \label{bkm:FigEBSDvsPF}
  \end{center}
\end{figure}

Figure~\ref{bkm:FigFracVsTime_1} shows the simulated evolution of $\alpha'$ and $\beta$ volume fractions
($f_{\alpha'}$ and $f_{\beta}$, respectively) during the step-wise heat treatment, and the corresponding thermal history (PF input data). 
The microstructure evolution in regions 1 and 2, at the end of the 8 plateaus of the annealing process ($t=t_{1}$ to $t_{8}$ labeled in Figure~\ref{bkm:FigFracVsTime_1}) is shown in Figures~\ref{bkm:FigPFregion1} and \ref{bkm:FigPFregion2}. 
The solute (V) concentration map in both regions at the end of the 700$^{\circ}$C ($t=t_3$) and 800$^{\circ}$C ($t=t_4$) plateaus, is illustrated in Figure~\ref{bkm:FigPFregions1-2_conc}.
The concentration $c$, and corresponding $\eta_i$, along a scanning line shown in Figure~\ref{bkm:FigPFregion1} (region 1), is presented in Figure~\ref{bkm:V_concentration_line}.
Stable $\alpha$ and metastable $\alpha'$ phases may be challenging to distinguish, since they share a similar crystal structure and close lattice parameters \cite{bkm:RefTan2016-31}. 
The main differences appear in their morphologies and the solute concentration, with $\alpha'$ typically exhibiting a fine lath structure and solute supersaturation beyond equilibrium.
Yet, the supersaturation threshold to unambiguously determine $\alpha$ and $\alpha'$ regions  is not clearly established.
Here, in order to distinguish $\alpha$ and $\alpha'$ phases, we use an {\it ad hoc} criterion and consider that a point within the $\alpha/\alpha'$ region (i.e. where $\sum_i\eta_i>0.99$) belongs to the metastable $\alpha'$ phase if its concentration is higher than $1.5c_{\alpha}^{*}$, where $c_{\alpha}^{*}$ is the V equilibrium concentration in the $\alpha$ phase at a given temperature. 
Finally, Figure~\ref{bkm:FigEBSDvsPF} compares experimental (EBSD) maps and PF-simulated microstructures after 300\,s at $T=850^{\circ}$C (i.e. $t_4<t<t_5$) for regions 1 and 2. 

%
\section{Discussion}
%

\subsection{Experiments}
\label{bkm:Ref122788851}

The microstructure in as-built L-PBF Ti-6Al-4V (Figure~\ref{bkm:FigExptsICs}), composed of fully acicular $\alpha'$ martensite, no detectable $\beta$ phase, and a weak crystallographic texture, is quite typical for as-built L-PBF Ti-6Al-4V microstructures, and consistent with previously reported results \cite{bkm:RefZhang2018-36,bkm:RefSabban2019-70}.
The formation of the $\alpha'$ martensite in the as-built microstructure (Figure~\ref{bkm:FigExptsICs}) can be attributed to the fast cooling of the melt pool during L-PBF. 
High-speed thermal imaging measurements suggest that the effective cooling rate experienced in the L-PBF
Ti-6Al-4V melt pool is of order $10^6$\,$^{\circ}$C/s \cite{bkm:RefHooper2018-94}, while the \textit{in-situ} XRD analysis suggests an
overall cooling rate of around $10^4$\,$^{\circ}$C/s \cite{bkm:RefCalta2020-72}, both being far beyond the critical cooling
rate required to form metastable martensite upon cooling (410$^{\circ}$C/s) \cite{bkm:RefAhmed1998-95}. Recent research
suggests that, under such high cooling rates, nano-scale $\beta$ stripes may be preserved in the as-built microstructure alongside
the dominant $\alpha'$ martensite \cite{bkm:RefZafari2018-21}. However, such nano-scale $\beta$ phases are too small
to be detected by SEM or XRD \cite{bkm:RefZafari2018-21}.

During the thermal treatment, the microstructure does not exhibit any visible change at 400$^{\circ}$C and 500$^{\circ}$C (Figure~\ref{bkm:FigExptsVsTime}(a)-(b)). 
At this low temperature, recovery of the sub-grain structure occurs, which includes relaxation of the residual stress, annihilation and rearrangement of the
dislocations \cite{bkm:RefCallister2011-96}. Such a recovery process was evidenced recently, via \textit{in-situ} high-energy XRD,
by the reduction of XRD peak width (Full Width at Half Maximum, FWHM) in all observed grain orientations at 497$^{\circ}$C
\cite{bkm:RefBrown2021-97}. This recovery occurs without the migration of any high-angle grain boundaries and will not
trigger any evolution in the microstructural constituents \cite{bkm:RefSallicaLeva2016-64,bkm:RefCallister2011-96}. Therefore,
no visible changes can be noticed in the forescatter electron micrographs.

Visible changes can be observed at higher temperatures. Coarsening of the microstructure starts being noticeable at
600$^{\circ}$C (Figure~\ref{bkm:FigExptsVsTime}(c)), leading to the emergence of topographical changes. 
At 700$^{\circ}$C (Figure~\ref{bkm:FigExptsVsTime}(d)), the grain structure appears clearly, and the boundary of each $\alpha$/$\alpha'$
grain is well marked. Solute diffusion and elements partitioning become significant at elevated temperature, which
leads to the decomposition of $\alpha'$ into $\alpha+\beta$ \cite{bkm:RefZhang2018-36,bkm:RefSallicaLeva2016-64}. 
Visible changes in the topography of the grain boundary are attributed to the nucleation and growth of the $\beta$ phase,
which takes place on the boundary of the newly formed $\alpha$ \cite{bkm:RefZhang2018-36}. Previous studies reported
martensite start temperatures $M_{S}^*$ ranging from 580$^{\circ}$C to 700$^{\circ}$C, affected by multiple
factors, such as the dislocation density, residual stress, and supersaturation in V
\cite{bkm:RefYang2016-28,bkm:RefQazi2003-98,bkm:RefNeelakantan2009-99,bkm:RefChong2017-100}. The initial appearance of the $\beta$ phase was previously captured by \textit{in-situ} high-energy XRD analysis of L-PBF Ti-6Al-4V at 640$^{\circ}$C \cite{bkm:RefBrown2021-97}, which supports our \textit{in-situ} observations.

At higher temperature, e.g., 800$^{\circ}$C (Figure~\ref{bkm:FigExptsVsTime}(e)), we expect the original $\alpha'$ to be nearly fully decomposed
\cite{bkm:RefZhang2018-36} and we observed a further coarsening of the $\alpha$ phase. 
A notable growth of the $\beta$ phase can be observed at 850$^{\circ}$C as seen in the measured $\beta$ orientation
map (Figure~\ref{bkm:FigExptsVsTime}(i)). At this temperature, regions that consist of $\beta$ phases appear with a specific stepped
topography in the secondary electron micrographs (Figure~\ref{bkm:FigExptsVsTime}(f)). This topography might result from the bcc structure of the $\beta$ phase. 
The $\beta$ phase can be observed both at the $\alpha$ boundary and in the interior of the $\alpha$
phase, as indicated by the white arrow in Figure~\ref{bkm:FigExptsVsTime}(i). 
The observed inner $\beta$ regions might stem from nucleation at lattice defects located within the $\alpha$/$\alpha'$ phase.
This is consistent with recent studies suggesting that, beyond nucleation at vanadium-enriched grain/lath boundaries, the $\beta$ phase can also nucleate along lattice defects within $\alpha$/$\alpha'$ phases 
\cite{bkm:RefTan2016-31,bkm:RefHaubrich2019-22,bkm:RefZou2020-51}. 
The most likely underlying causes of such nucleation events are (i) the high density of lattice defects (e.g. dislocations, stacking faults, twin boundaries) reported to form in Ti-6Al-4V processed at high cooling rates \cite{bkm:RefXu2015-11, bkm:RefYang2016-28, bkm:RefWu2016-29, bkm:RefZhang2018-36}, and (ii) the local V enrichment at one-dimensional lattice defects such as dislocations and dissociated (partial) dislocations, acting as precursor nucleation sites for the $\beta$ phase \cite{bkm:RefTan2016-31, bkm:RefHaubrich2019-22}.
The proportion of $\beta$-phase nucleating within the bulk $\alpha$/$\alpha'$ phase remains low compared to that nucleation along grain/lath boundaries (Figure~\ref{bkm:FigExptsVsTime}(i)).
Therefore, in spite of the rudimentary description of nucleation in the present model, we expect the resulting discrepancies to be relatively small.

Recrystallised high-angle boundary (HAB) $\beta$ phase can also be observed in the orientation map, as indicated by the black arrows in Figure~\ref{bkm:FigExptsVsTime}(i). 
This recrystallisation process cannot usually be produced in traditional manufactured Ti-6Al-4V via simple heating
\cite{bkm:RefIvasishin1999-101}. Therefore, it is a unique feature of the additively manufactured Ti-6Al-4V. Recent research suggests that
this recrystallisation may be triggered by the stored energy of deformation in the as-built microstructure, which is typically
expressed by the large dislocation densities found after L-PBF \cite{bkm:RefZou2020-51}.

The microstructure at 900$^{\circ}$C is shown in Figure~\ref{bkm:FigExptsVsTime}(g). An increase of the area covered by stepped topography can
be observed when compared to a lower temperature, which evidences the increase of $\beta$ phase in the $\alpha+\beta$ matrix, as the temperature
approaches the $\beta$ transus temperature. However, the prior-$\beta$ grain boundary is nearly intact and unchanged at 900$^{\circ}$C (see red dotted line in Figure~\ref{bkm:FigExptsVsTime}(g) marking the location of the initial prior-$\beta$ GB). 
It is consistent with results from other studies, which reported that heat treatment at a temperature below $\beta$ transus has limited influence on the prior-$\beta$ structure of L-PBF Ti-6Al-4V \cite{bkm:RefZhang2018-36,bkm:RefSabban2019-70}.

\subsection{Model validations}
\label{sec:resu:validation}

The comparison of PF-predicted, theoretical (lever rule), and experimentally assessed $\beta$ volume fraction (Figure~\ref{bkm:FigLEVERvsPFvsExpts}) serves as a validation of our model and parameters for thermodynamic equilibrium. 
As expected, a temperature increase clearly reduces the incubation time and increases the transformation rate. 
Notably, the $f_\beta(t)$ curves slightly deviate from the smooth sigmoidal shape characteristic of Avrami kinetics, due to the additional effect of the microstructure (e.g., accounting for interfacial energies and kinetics), particularly at high temperatures undergoing a relatively more complex interplay of solute diffusion and interface kinetics.

The different thermal paths (Figure~\ref{bkm:gilmur}) influence mostly the early onset of the transformation, but they nonetheless result in clearly distinct kinetics, since transformation kinetics (e.g. Avrami plots) are usually represented in logarithmic scale. 
Accounting for the heating up of the sample leads, as expected, to a slower transformation, and to a good agreement with experiments.
This validates that the values considered for mobilities $M$ in Eq.~\eqref{eq:cahn-hilliard}
(see Appendix~\ref{appendix:atomicm}) and $L$ in Eq.~\eqref{eq:allen-cahn} (see Sec.~\ref{sec:parameters}) are reasonable. We
could have achieved a better match between experiments and simulations by further adjusting the mobilities, but, given the degree
of uncertainties associated with the experimental data, we chose to keep the atomic mobilities as developed in
Ref.~\cite{bkm:RefGierlotka2019-93} (see Appendix~\ref{appendix:atomicm}) and to not modify the interface mobility $L$ any
further.
Notably, in Figure~\ref{bkm:gilmur}, the agreement between PF results and experimental measurements is better for the 700$^\circ$C case than for the 800$^\circ$C case, which we attribute to differences in $\alpha'$ initial microstructures. 
The experimental data relates to wrought and water-quenched samples and, as mentioned in \cite{bkm:RefVilaro2011-10}, coarser $\alpha'$ martensite laths are generally obtained in comparison with L-PBF processed samples. 
Therefore, fewer nucleation sites are expected for the $\beta$ phase, resulting in a slower kinetics of the transformation, even more marked at high temperatures where $\beta$ nuclei grow more rapidly.

\subsection{Martensite decomposition in post-AM heat treatments}

The temperature plateau at 400$^{\circ}$C does not induce any noticeable transformation (Figure~\ref{bkm:FigFracVsTime_1}), either $(\alpha'\rightarrow\alpha)$ or $(\alpha'\rightarrow\alpha+\beta)$. 
At 500$^{\circ}$C, there still is no significant sign of $\beta$ nucleation (i.e. $\sum_i\eta_i$ remains close to 1), but the diffusion of V is now fast enough to initiate the transformation of $\alpha'$ into $\alpha$ (see Figure~\ref{bkm:FigFracVsTime_1}).
Higher temperatures ($T\geq600^\circ$C) lead to the formation of $\beta$, first nucleating along $\alpha'$ martensite lath boundaries at $T=600^{\circ}$C (white arrows in Figs~\ref{bkm:FigPFregion1}b and \ref{bkm:FigPFregion2}b), then growing as thin layers between the laths at $T=700^{\circ}$C and 800$^{\circ}$C
(panels c and d), as reported in various studies \cite{bkm:RefTerHaar2018-13,bkm:RefZeng2005-62,bkm:RefSallicaLeva2016-64,bkm:RefGupta2016-66,bkm:RefKaschel2020b-67}.
At $T=850^{\circ}$C (Figs~\ref{bkm:FigPFregion1}e and \ref{bkm:FigPFregion2}e), remaining $\alpha$ regions, now fully transformed from $\alpha'$, are still occupying nearly 70\% of the domain (see Figure~\ref{bkm:FigFracVsTime_1}), and are fully surrounded by $\beta$ layers.
As the temperature increases further above 850$^{\circ}$C (Figs~\ref{bkm:FigPFregion1}f-h and \ref{bkm:FigPFregion2}f-h), the remaining $\alpha$ progressively transforms into $\beta$, as diffusion is now sufficiently fast to reach near-equilibrium phase fractions within the isothermal plateaus.

The evolution of the $\beta$ fraction (Figure~\ref{bkm:FigFracVsTime_1}) exhibits multiple sigmoid-like plateaus as the annealing temperature evolves. 
At 600$^{\circ}$C the $\beta$ volume fraction ($f_{\beta}<0.015$) remains far from equilibrium ($f_{\beta}^{*}\approx0.09$) due to the low mobilities of V and interfaces at such low temperature and to the short annealing times. 
Once at $T\geq700^{\circ}$C, the V concentration in $\beta$ is close to equilibrium ($c_{\beta}^{*}$), which decreases as the temperature increases (see Figure~\ref{bkm:V_concentration_line}).
However, at 900$^{\circ}$C and 980$^{\circ}$C, the isothermal plateaus are too short to reach full equilibrium, which can be seen in the still significant slope of the $f_\beta(t)$ curve at $t_{6}$ and $t_{8}$, indicated with red arrows in Figure~\ref{bkm:FigFracVsTime_1}. 
As expected, the time derivative of $f_\beta$ indicates that the transformation rate increases with temperature, for the most part due to the increase of V mobility and interface mobility. 
The transformation rate in region 1, initially slower at low $T$, becomes higher than that of region 2 above 900$^{\circ}$C, which we attribute to the initially finer $\alpha'$ martensitic microstructure of region 1 (Figure~\ref{bkm:FigICs}).

The V concentration along the scanning line of region 1 (Figure~\ref{bkm:V_concentration_line}) shows that the stable $\alpha$ and martensitic $\alpha'$ phases coexist at 600$^{\circ}$C and 700$^{\circ}$C (with $c_{\alpha'}$ labelled with symbols when $c>1.5c_\alpha^*$).
 At 700$^{\circ}$C, thin $\alpha'$ laths have been completely converted to $\alpha$ but the thicker ones are only partially transformed. 
This appears clearly in Figure~\ref{bkm:FigPFregions1-2_conc}, as well as Figures~\ref{bkm:FigPFregion1} and \ref{bkm:FigPFregion2} (panels c), where we applied a darker colour shade in regions corresponding to $\alpha'$ (i.e. where $c>1.5c_\alpha^*$).
This partial $\alpha'\rightarrow\alpha$ transformation is attributed to the limited annealing times, thus not allowing the diffusion of V across the entire thicker laths, given the moderate diffusivity at this temperature. 
Indeed, since the $\alpha'\rightarrow\alpha$ transformation has already started at $T=600^\circ$C (see Figures~\ref{bkm:FigFracVsTime_1} and \ref{bkm:V_concentration_line}), one would expect it to be completed provided a sufficient annealing time.
Finally, at $T\geq800^{\circ}$C, thicker laths are completely transformed into $\alpha$ and their V composition remains very close to equilibrium ($c_{\alpha}^{*}$). 

Importantly, our simulations confirm that, for the considered annealing times, the microstructure conserves a (ultra)fine $\alpha+\beta$ microstructure inherited from the initial $\alpha'$ lath structures (Figure~\ref{bkm:FigICs}) at 850$^{\circ}$C, while nearly all the $\alpha'$ martensite has transformed into an equilibrium $\alpha$ phase from $T=800^\circ$C.
By increasing the annealing temperature above 850$^\circ$C, small $\alpha$ grains transform into $\beta$ and larger $\alpha$ grains undergo further coarsening (hinting at an evolution toward a globularisation regime).
These results are all consistent with previous experiments showing that below $400^\circ$C only stress relaxation occurs \cite{bkm:RefKaschel2020b-67}, while phase transformation takes place above $550^\circ$C \cite{bkm:RefKaschel2020b-67}, but that 2h of annealing time may still not be sufficient to complete the $\alpha'\rightarrow\alpha+\beta$ transformation even at $700^\circ$C or $800^\circ$C \cite{bkm:RefCao2018-35}, while microstructural coarsening is only evident at $900^\circ$C and above \cite{bkm:RefZhang2018-36}, and with globularisation occurring for heat treatments usually involving temperatures above $800^\circ$C (most often above $900^\circ$C) \cite{bkm:RefSabban2019-70,xiao2022mechanism,semiatin2005prediction,stefansson2002kinetics}.

Finally, regarding the comparison between experimental and simulated microstructures at $T=850^{\circ}$C (Figure~\ref{bkm:FigEBSDvsPF}), our simulations capture well the fact that thinner laths were transformed into $\beta$ phase and that $\alpha$ conserved a lath shape inherited from the $\alpha'$ martensite. 
As already apparent in Figure~\ref{bkm:FigLEVERvsPFvsExpts}, the volume fractions of $\beta$ phase are in reasonable agreement between simulation (region~1: $f_{\beta}=0.289$, region~2: $f_{\beta}=0.281$) and experiments (region~1: $f_{\beta}=0.26$, region~2: $f_{\beta}=0.18$), which is relatively close to equilibrium ($f_{\beta}^{*}\approx0.31$).
The discrepancy in region 2 is attributed primarily to dimensionality, since the experimental $f_\beta$ is measured from a single 2D EBSD slice and simulations are also two-dimensional. 
Another difference observed in Figure~\ref{bkm:FigEBSDvsPF} is the propensity of the PF model to grow $\beta$ phase homogeneously along all $\alpha'$ lath boundaries, hence resulting in a $\beta$ layer fully surrounding the remaining $\alpha$ grains, while the EBSD maps still contains grain boundaries between different $\alpha$ grains.
In addition to the already-mentioned dimensionality, here we also attribute this discrepancy to the rudimentary treatment of $\beta$ nucleation in the model -- or, more precisely, the lack of a specific nucleation algorithm, as we let $\beta$ nucleate spontaneously at grain boundaries, triggered by fluctuations from the interpenetration term in the free energy (Eq.~\eqref{eq:g}).
As a result, simulations tend to predict a continuous nucleation of $\beta$ along the martensitic lath boundaries, ultimately resulting in a relatively finer microstructure than observed in experiments.

%
\section{Summary, Conclusions and Perspectives}
%

We performed a joint experimental-computational study of martensite decomposition during post-printing heat treatment of
L-PBF additively manufactured Ti-6Al-4V alloy. Our original experiments made use of \textit{in-situ} electron microscopy and
diffraction (EBSD) analysis during heat treatment up to nearly $\beta$-transus temperature. Our simulations rely on a FFT-based
GPU-accelerated phase-field model of microstructure evolution, using experimental ESBD maps of as-built Ti-6Al-4V and
experimental temperature history as input.

Our \textit{in-situ} experiments confirmed or revealed that:
\begin{itemize}
\item The microstructure of L-PBF processed Ti-6Al-4V alloy is fully martensitic ($\alpha'$),
  with a weak texture mostly due to the several possible lath orientations emerging from prior-$\beta$ grains (up to 12
  variants per $\beta$ grain). Possible other phases, such as small $\beta$ grains
  \cite{bkm:RefZafari2018-21,bkm:RefHaubrich2019-22} or precipitates \cite{bkm:RefThijs2010-23} were not observed -- but
  they might have been too small to unambiguously identify with the techniques employed here (SEM, EBSD, XRD).
\item At low annealing temperature ($T=400^{\circ}$C or 500$^{\circ}$C), while some relaxation of residual stresses or
  annihilation/rearrangement of dislocations might have occurred, the topology of the microstructure did not exhibit any
  significant change. Visible microstructure evolution occurred only at $T\geq700^{\circ}$C. 
\item Nucleation of $\beta$ phase primarily takes place along the boundaries of $\alpha'$ laths. However, 
  traces of $\beta$ phase nucleating within $\alpha/\alpha'$ region, rather than at lath boundaries, suggests possible
  $\beta$ nucleation sites along lattice defects (e.g. dislocations, twins, sub-grain boundaries, etc.)
  \cite{bkm:RefHaubrich2019-22,bkm:RefZou2020-51}.
\item While it substantially alters the $\alpha'$ lath structure within the prior-$\beta$ grain, a near
  $\beta$-transus temperature does not seem to significantly change the topology of the $\beta$ grain structure.
\end{itemize}

From the modelling and simulations perspective:
\begin{itemize}
\item We proposed a phase-field model, kept relatively simple by the adoption of experiment/evidence-motivated assumptions, in particular considering (i) a pseudo-binary alloy approximation, (ii) isotropic interfaces, and
  (iii) 2D simulations.
  Thermodynamic description of phases (free energies and mobilities) were directly taken from the literature (with almost no adjusted parameters).
  Simulations were accelerated by an original FFT-based resolution algorithm and GPU parallelisation. 
  We used the model to simulate the evolution of experimentally characterised microstructures (using EBSD maps as initial conditions) and experimentally imposed/measured thermal history.
\item Despite its simplicity, the model captures the salient features of the $\alpha' \to \alpha+\beta$
  transformation, such as: (i) a martensite decomposition visibly starting around
  600$^{\circ}$C with a small amount of $\beta$ phase formed along $\alpha'$ grain boundaries, (ii) a transformation
  kinetics leading to near-equilibrium fractions of $\alpha$ and $\beta$ phases at the end of the isothermal plateaus
  at $T \geq700^{\circ}$C, (iii) a reasonable direct comparison of EBSD maps at $T=850^{\circ}$C (Figure~\ref{bkm:FigEBSDvsPF}) -- however with slightly finer predicted microstructure, attributed to the primitive modeling of $\beta$ nucleation (or, more precisely, a lack thereof).
\item Our simulations complemented the \textit{in-situ} experimental characterisation, in particular by providing the
  spatiotemporal evolution of solute (V) within the inspected regions. This allowed, for instance, to ascertain that, for
  the considered temperature-time profile, the full decomposition of $\alpha'$ into an $\alpha+\beta$ structure
  was complete at $T\approx800^{\circ}$C.
\end{itemize}

This work emphasizes the advantage of relatively simple phase-field models, which provide a deeper description of phase transformation kinetics than classical mean-field (e.g. Avrami-based) models commonly employed to analyse martensite decomposition in additively manufactured Ti-6Al-4V \cite{bkm:RefGilMur1996-73,bkm:RefMurgau2012-74,bkm:RefYang2021-75}.
Indeed, PF models provide not only fractions of phase, but also the morphological evolution of the microstructure.
Such information may be critical, for instance in the production of ultrafine $\alpha+\beta$ microstructures with
outstanding mechanical properties. 

Limitations of the simulations carried out here pertain most importantly to the dimensionality of the simulations (2D)
and to some fairly restrictive model assumptions, in particular neglecting the effects of mechanics (e.g. strain energy contributions) and anisotropy of bulk crystalline phases and interfaces. 
A more advanced treatment of $\beta$ nucleation \cite{granasy2019phase}, possibly accounting for the spatial distribution of defects (e.g. dislocation) and their effect on the local concentration fields and contribution to the nucleation energy barrier, could also allow to achieve a greater agreement with experiments.

Potential perspectives from this work are manifold, some of which are already the focus of work in progress.
The model could be used to assess whether and how a different solute concentration across $\alpha'$ laths (as evidenced recently \cite{zhao2023influence}) might affect the kinetics of martensite decomposition.
It could also be used to explore the effect of minor compositional changes \cite{du2023facile} or 
non-conventional heat treatments, e.g. cyclic heat treatments \cite{bkm:RefSabban2019-70,mckenna2023evaluation} or intrinsic heat cycling during the printing process \cite{bkm:RefXu2015-11,bkm:RefXu2016-12,bkm:RefSimonelli2014-27,bkm:RefBarrioberoVila2017-34}.
Our ongoing efforts focus on the incorporation of anisotropic elastic terms, in order to better account for the underlying crystal structure and grain orientations.
Such extension, combined with further thermodynamic description of possible metastable phases, could for instance allow studying in greater detail the still debated reaction path between $\alpha'$ and $\alpha+\beta$, e.g. addressing the potential intermediate formation of a non-equilibrium $\alpha$-like phase with a hcp structure but a composition close to that of the $\beta$ phase \cite{wang2022formation}.

By including a more proper description of mechanical effects, an underlying objective is to ultimately use microstructures predicted from virtual processing into a micromechanics framework also based on FFT spectral methods \cite{bkm:RefLucarini2019-104} in order to simulate their mechanical behaviour and predict their mechanical properties. 
Furthermore, while here we chose to use experimental martensitic microstructures as initial conditions, the proposed model could greatly benefit from a coupling (upstream) with models capable of predicting the morphology of martensitic microstructures emerging from L-PBF of Ti-6Al-4V (e.g. \cite{xiang2023phase}) and (downstream) with models capable to assess the mechanical properties of resulting heterogeneous microstructure (e.g. \cite{shi2019integrated, liu2020integration}).
Our ambition is that such an integrated predictive framework could lead to the design of original thermal treatments to maximise the mechanical properties of L-PBF Ti-6Al-4V, but also to the design of novel Ti-based alloys.
Ultimately, this concerted effort from the multiscale modeling community shall allow to take full advantage of additive
manufacturing and post processing routes, and accelerate the synergistic development of novel alloys and manufacturing technologies, e.g. in the aeronautical and biomedical sectors.


\section*{CRediT Author Contribution Statement}
{\bf A.D.~Boccardo}: Investigation (Computational); Methodology; Formal analysis; Software; Validation; Funding acquisition; Writing - original draft.
{\bf Z.~Zou}: Investigation (Experimental); Methodology; Formal analysis; Writing - original draft.
{\bf M.~Simonelli}: Investigation (Experimental); Methodology; Formal analysis; Funding acquisition; Supervision; 
{\bf M.~Tong}: Conceptualization; Supervision; 
{\bf J.~Segurado}: Conceptualization; Supervision; 
{\bf S.B.~Leen}: Conceptualization; Funding acquisition; Supervision; 
{\bf D.~Tourret}: Conceptualization; Methodology; Funding acquisition; Supervision; Writing - original draft.
{\bf All}: Writing - Review \& Editing.

%
\section*{Data availability}
%

The data required to reproduce these findings cannot be shared at this time as the data also forms part of an ongoing study. However, any shareable piece of source code, file, or post processing script will be gladly shared upon request to the corresponding author.

%
\section*{Acknowledgements}
%

This publication has emanated from research supported in part by a grant from Science Foundation Ireland under
Grant number 16/RC/3872. For the purpose of Open Access, the author has applied a CC BY public copyright licence
to any Author Accepted Manuscript version arising from this submission. ADB acknowledges the financial support from
the European Commission through the M3TiAM project (HORIZON-TMA-MSCA-PF-EF 2021, Grant agreement 101063099). 
DT gratefully acknowledges support from the Spanish Ministry of Science through a Ram\'on y Cajal
Fellowship (RYC2019-028233-I).
Experimental work has been made possible by funding provided through the University of Nottingham's Nottingham
Research Fellowship. Thanks to Dr. Nigel Neate (University of Nottingham) for his assistance on the high temperature
microscopy. 

\appendix

%
\section{Redlich-Kister Polynomials}
%

\subsection{Chemical free energy}
\label{appendix:chemfe}

Chemical potentials of the alloying elements in $\alpha$ and $\beta$ phases were calculated at different
temperatures in the range from 400$^{\circ}$C to 1000$^{\circ}$C using ThermoCalc (database: TCNI8) and then approximated by cubic polynomials:

\begin{align}
  \mu_{Ti}^{\alpha}=~&6.5215\times 10^{-6}\,T^3-3.6174\times 10^{-2}\,T^2 
  -12.930\,T-2.1944\times 10^3
  \\
  \mu_{V}^{\alpha}=~&6.5307\times 10^{-6}\,T^3-3.4676\times 10^{-2}\,T^2 
  -11.591\,T-1.9335\times 10^3
  \\
  \mu_{Al}^{\alpha}=~&6.4497\times 10^{-6}\,T^3-3.5642\times 10^{-2}\,T^2 
  -12.247\,T-3.2008\times 10^3
\end{align}
\begin{align}
  \mu_{Ti}^{\beta}=~&6.8957\times 10^{-6}\,T^3-3.5142\times 10^{-2}\,T^2 
  -20.431\,T-4.5117\times 10^3
  \\
  \mu_{V}^{\beta}=~&6.5306\times 10^{-6}\,T^3-3.4675\times 10^{-2}\,T^2 
  -13.991\,T-2.0665\times 10^3
  \\
  \mu_{Al}^{\beta}=~&6.4407\times 10^{-6}\,T^3-3.5627\times 10^{-2}\,T^2 
  -15.268\,T-7.8040\times 10^3
\end{align}

\noindent
Redlich-Kister coefficients of the excess term in $\alpha$ and $\beta$ phases, extracted from
Ref.~\cite{bkm:RefAnsara1998-90}, except for $L_{\mathit{AlTi}_0}^{\beta}$ that was recalibrated (see Section~\ref{sec:meth:pf}), are:

\begin{align}
  &L_{\mathit{TiV}_0}^{\alpha}=20000
  \\
  &L_{\mathit{AlTi}_0}^{\alpha }=-133500+39\,T
  \\
  &L_{\mathit{AlTi}_1}^{\alpha }=750
  \\
  &L_{\mathit{AlTi}_2}^{\alpha }=17500
\end{align}
\begin{align}
  &L_{\mathit{TiV}_0}^{\beta}=10500-1.5\,T
  \\
  &L_{\mathit{TiV}_1}^{\beta}=2000
  \\
  &L_{\mathit{TiV}_2}^{\beta}=1000
  \\
  &L_{\mathit{AlTi}_0}^{\beta}=-118500+33.5\,T
  \\
  &L_{\mathit{AlTi}_1}^{\beta}=6000
  \\
  &L_{\mathit{AlTi}_2}^{\beta}=21200
  \\
  &L_{\mathit{AlV}_0}^{\beta}=-95000+20\,T
  \\
  &L_{\mathit{AlV}_1}^{\beta}=-6000
\end{align}

\subsection{Atomic mobility}
\label{appendix:atomicm}

Atomic mobility of alloying elements in phase $\varphi\in(\alpha,\beta)$, extracted from Ref.~\cite{bkm:RefGierlotka2019-93},
are:

\begin{align}
  M_{Ti}^{\varphi}=\exp \bigg{[} \dfrac{\Delta G_{Ti}^{\varphi}}{RT} \bigg{]} \dfrac{1}{RT}
  ~;\quad
  M_{V}^{\varphi}=\exp \bigg{[} \dfrac{\Delta G_{V}^{\varphi}}{RT} \bigg{]}\dfrac{1}{RT}
  ~;\quad
  M_{Al}^{\varphi}=\exp \bigg{[} \dfrac{\Delta G_{Al}^{\varphi}}{RT} \bigg{]} \dfrac{1}{RT}
\end{align}

\noindent where:

\begin{align}
  &\Delta G_{Ti}^{\alpha}=c_{Ti}\Delta G_{Ti,Ti}^{\alpha}+c_{V}\Delta G_{Ti,V}^{\alpha}+c_{Al}\Delta G_{Ti,Al}^{\alpha}
  \\
  &\Delta G_{Ti,Ti}^{\alpha}=-291755.355-115.51861\,T
  \\
  &\Delta G_{Ti,V}^{\alpha}=-298988.939-333.55846\,T
  \\
  &\Delta G_{Ti,Al}^{\alpha}=-111539.54-141.61855\,T
\end{align}
\begin{align}
  &\Delta G_{V}^{\alpha}=c_{Ti}\Delta G_{V,Ti}^{\alpha}+c_{V}\Delta G_{V,V}^{\alpha}+c_{Al}\Delta G_{V,Al}^{\alpha}
  \\
  &\Delta G_{V,Ti}^{\alpha}=-262867.416-95.5291479\,T
  \\
  &\Delta G_{V,V}^{\alpha}=-270101.0-313.569\,T
  \\
  &\Delta G_{V,Al}^{\alpha}=0
\end{align}
\begin{align}
  &\Delta G_{Al}^{\alpha}=c_{Ti}\Delta G_{Al,Ti}^{\alpha}+c_{V}\Delta G_{Al,V}^{\alpha}+c_{Al}\Delta G_{Al,Al}^{\alpha}
  \\
  &\Delta G_{Al,Ti}^{\alpha}=-311534.815-95.17706\,T
  \\
  &\Delta G_{Al,V}^{\alpha}=0
  \\
  &\Delta G_{Al,Al}^{\alpha}=-131319.0-121.277\,T
\end{align}

\noindent and:

\begin{align}
  &\Delta G_{Ti}^{\beta}=c_{Ti}\mathit{\Delta G}_{Ti,Ti}^{\beta}+c_{V}\Delta G_{Ti,V}^{\beta}+c_{Al}\Delta G_{Ti,Al}^{\beta}+c_{Al}c_{Ti}\Delta G_{Ti,Al,Ti}^{\beta} \nonumber \\
  & \ \ \ \ \ \ \ \ \ \ \ +c_{Ti}c_{V}\big{[}\Delta G_{Ti,Ti,V_{0}}^{\beta}+\Delta G_{Ti,Ti,V_{1}}^{\beta}(c_{Ti}-c_{V})\big{]}
  \\
  &\Delta G_{Ti,Al}^{\beta}=-104715.111-267.14\,T
  \\
  &\Delta G_{Ti,Ti}^{\beta}=-153362.077-126.135031\,T
  \\
  &\Delta G_{Ti,V}^{\beta}=-331128.104-65.0814985\,T
  \\
  &\Delta G_{Ti,Al,Ti}^{\beta}=-139719.202+294.400646\,T
  \\
  &\Delta G_{Ti,Ti,V_{0}}^{\beta}=234932.489-219.426462\,T
  \\
  &\Delta G_{Ti,Ti,V_{1}}^{\beta}=-261099.972
\end{align}
\begin{align}
  &\Delta G_{V}^{\beta}=c_{Ti}\mathit{\Delta G}_{V,Ti}^{\beta}+c_{V}\Delta G_{V,V}^{\beta}+c_{Al}\Delta G_{V,Al}^{\beta}+c_{Al}c_{Ti}\Delta G_{V,Al,Ti}^{\beta} \nonumber \\
  & \ \ \ \ \ \ \ \ \ \ \ +c_{Ti}c_{V}\big{[}\Delta G_{V,Ti,V_{0}}^{\beta}+\Delta G_{V,Ti,V_{1}}^{\beta}(c_{Ti}-c_{V})\big{]}
  \\
  &\Delta G_{V,Ti}^{\beta}=-179114.192-107.062681\,T
  \\
  &\Delta G_{V,Al}^{\beta}=0
  \\
  &\Delta G_{V,V}^{\beta}=-322244.697-75.7262222\,T
  \\
  &\Delta G_{V,Al,Ti}^{\beta}=421533.074-629.054528\,T
  \\
  &\Delta G_{V,Ti,V_{0}}^{\beta}=163888.616-38.4124933\,T
  \\
  &\Delta G_{V,Ti,V_{1}}^{\beta}=-54096.709
\end{align}
\begin{align}
  &\Delta G_{Al}^{\beta}=c_{Ti}\Delta G_{Al,Ti}^{\beta}+c_{V}\Delta G_{Al,V}^{\beta}+c_{Al}\Delta G_{Al,Al}^{\beta}+c_{Al}c_{Ti}\Delta G_{Al,Al,Ti}^{\beta} \nonumber \\
  & \ \ \ \ \ \ \ \ \ \ \ +c_{Ti}c_{V}\Delta G_{Al,Ti,V}^{\beta}
  \\
  &\Delta G_{Al,Ti}^{\beta}=-159367.966-125.405247\,T
  \\
  &\Delta G_{Al,Al}^{\beta}=-110721.0-266.41\,T
  \\
  &\Delta G_{Al,V}^{\beta}=0
  \\
  &\Delta G_{Al,Al,Ti}^{\beta}=-348892.407+280.179819\,T
  \\
  &\Delta G_{Al,Ti,V}^{\beta}=-261912.01-157.174295\,T
\end{align}

%
\section{Numerical Methods}
\label{appendix:numerics}
%

\subsection{Cahn-Hilliard equation}
\label{appendix:CH}

After applying a backward Euler implicit time discretisation, Eq.~\eqref{eq:cahn-hilliard} is written as:

\begin{align}
  \big{(}c^{(t+\Delta t)}-c^{(t)}\big{)}-V_{m}^{2}\Delta t \nabla \cdot \bigg{[}\dfrac{M^{(t+\Delta t)}}{V_{m}}\nabla\bigg{(}\dfrac{\partial f_c^{(t+\Delta t)}}{\partial c} -\kappa_c\nabla^{2}c^{(t+\Delta t)} \bigg{)}\bigg{]} = 0
  \label{eq:C-H_discrete}
\end{align}

For the sake of clarity, the unknown field $c^{(t+\Delta t)}$ will be referred to as $c$. The differential equation above is non-linear and will be solved iteratively by successive linearisations using the Newton method. To this aim, the concentration field at
iteration $(j+1)$ is written as $c_{(j+1)}=c_{(j)}+\delta c$, where $c_{(j)}$ is the concentration at previous iteration $j$, and
$\delta c$ is the increment of the concentration in this new iteration, now used as the unknown of the problem.

Evaluating Eq.~\eqref{eq:C-H_discrete} at iteration $(j+1)$ and linearising 

\begin{align}
M_{c_{(j+1)}}=M_{(c_{(j)}+\delta c)} \approx M_{c_{(j)}}+\delta c \frac{\mathrm{d} M_{c_{(j)}}}{\mathrm{d} c}
\end{align}

\noindent and

\begin{align}
\frac{\partial f_{c_{(j+1)}}}{\partial c}=\frac{\partial f_{c (c_{(j)}+\delta c)}}{\partial c} \approx \frac{\partial f_{c_{(j)}}}{\partial c}+\delta c\frac{\partial^{2} f_{c_{(j)}}}{\partial c^{2}}
\end{align}

\noindent we obtain:

\begin{align}
  \big{(}c_{(j)}+\delta c-c^{(t)}\big{)}-V_{m}\Delta t \nabla \cdot \bigg{[} & \bigg{(}M_{c_{(j)}}+\dfrac{d M_{c_{(j)}}}{d c} \delta c\bigg{)} \nabla\bigg{(}\dfrac{\partial f_{c_{(j)}}}{\partial c} + \nonumber
  \\
  & \dfrac{\partial^{2} f_{c_{(j)}}}{\partial c^{2}}\delta c-\kappa_c\nabla^{2}(c_{(j)}+\delta c) \bigg{)}\bigg{]} = 0
  \label{eq:C-H_discrete_1}
\end{align}

Neglecting the high order terms related to $\delta c$, the following linear partial differential equation is obtained, whose solution
gives $\delta c$ that allows to compute $c_{(j+1)}$:

\begin{align}
  \delta c-V_{m}\Delta t \nabla \cdot \bigg{[} & M_{c_{(j)}} \nabla \bigg{(}\dfrac{\partial^{2} f_{c_{(j)}}}{\partial c^{2}}\delta c-\kappa_c\nabla^{2}\delta c \bigg{)}+ \nonumber
  \\
  & \dfrac{d M_{c_{(j)}}}{d c} \delta c \nabla \bigg{(}\dfrac{\partial f_{c_{(j)}}}{\partial c} - \kappa_c\nabla^{2}c_{(j)} \bigg{)} \bigg{]} = \nonumber
  \\
  & (c^{(t)}-c_{(j)})+ V_{m}\Delta t \nabla \cdot \bigg{[} M_{c_{(j)}} \nabla \bigg{(}\dfrac{\partial f_{c_{(j)}}}{\partial c}-\kappa_c\nabla^{2}c_{(j)} \bigg{)} \bigg{]} 
  \label{eq:C-H_discrete_2}
\end{align}

By definition of the Fourier transform, the gradient and Laplacian of a field $f$ are, respectively:

\begin{align}
  \widehat{\nabla f} &= i\boldsymbol{\xi} \widehat{f}
  \\
  \widehat{ \nabla^{2} f} &= -\| \boldsymbol{\xi} \|^{2} \widehat{f}
  \label{eq:grad}
\end{align}
with $i$ the imaginary unit, $\boldsymbol{\xi}$ the frequency vector, $\|\boldsymbol{\xi}\|^{2}$ the square of the frequency gradient,
and $\widehat{~}$ denotes the Fourier transform of the affected variable.

The previous differential equation can be transformed to the Fourier space, resulting in:

\begin{align}
  & \widehat{\delta c}-V_{m}\Delta t \ i \boldsymbol{\xi} \cdot \mathcal{F}\bigg{\{} M_{c_{(j)}} \mathcal{F}^{-1} \bigg{[} i \boldsymbol{\xi} \bigg{[}\mathcal{F}\bigg{(}\dfrac{\partial^{2} f_{c_{(j)}}}{\partial c^{2}}\mathcal{F}^{-1}\big{(}\widehat{\delta c}\big{)}\bigg{)}+\kappa_c \|\boldsymbol{\xi}\|^{2} \widehat{\delta c} \bigg{]} \bigg{]}+ \nonumber
  \\
  &\dfrac{d M_{c_{(j)}}}{d c} \mathcal{F}^{-1} \big{(}\widehat{\delta c}\big{)} \mathcal{F}^{-1} \bigg{[}i \boldsymbol{\xi} \bigg{(}\widehat{\dfrac{\partial f_{c _{(j)}}}{\partial c}} + \kappa_c \|\boldsymbol{\xi}\|^{2} \widehat{c}_{(j)} \bigg{)} \bigg{]} \bigg{\}} = (\widehat{c}^{(t)}-\widehat{c}_{(j)}) + \nonumber
  \\
  & V_{m}\Delta t \ i \boldsymbol{\xi} \cdot \mathcal{F} \bigg{\{} M_{c_{(j)}} \mathcal{F}^{-1} \bigg{[} i\boldsymbol{\xi} \bigg{(}\widehat{\dfrac{\partial f_{c_{(j)}}}{\partial c}}
    +\kappa_c \|\boldsymbol{\xi}\|^{2} \widehat{c}_{(j)} \bigg{)} \bigg{]} \bigg{\}}
  \label{eq:C-H_discrete_fft}
\end{align}

The resulting equation is linear and can be solved after discretisation using some algebraical linear solver combined with a preconditioner, as presented in Appendix \ref{appendix:resol}


\subsection{Allen-Cahn equation}
\label{appendix:AC}

Using a semi-implicit formulation, the ${\partial f_c}/{\partial \eta_{i}}$ terms are computed considering their value at the
previous time step (explicit part) and the Laplacian of $\eta_{i}$ is computed at the current time step (implicit
part). After applying time discretisation, Eq.~\eqref{eq:allen-cahn} is written as:
  
\begin{align}
  \big{(}\eta_{i}^{(t+\Delta t)}-\eta_{i}^{(t)}\big{)} +\dfrac{L}{V_{m}} \Delta t \bigg{(} \dfrac{\partial f_c^{(t)}}{\partial \eta_{i}} - \kappa_{\eta} \nabla^{2}\eta_{i}^{(t+\Delta t)} \bigg{)} = 0
  \label{eq:A-C_discrete}
\end{align}

For the sake of clarity, the unknown fields $\eta_{i}^{(t+\Delta t)}$ will be referred to as $\eta_{i}$.
The previous differential equations can be transformed to the Fourier space, resulting in:

\begin{align}
  \widehat{\eta}_{i}\bigg{(} 1+\dfrac{L}{V_{m}} \Delta t \kappa_{\eta} \|\boldsymbol{\xi}\|^2 \bigg{)} &= \widehat{\eta}_{i}^{(t)}-\dfrac{L}{V_{m}} \Delta t \widehat{\dfrac{\partial f_c^{(t)}}{\partial \eta_{i}}}
  \label{eq:A-C_discrete_fft}
\end{align}

At each time $t+\Delta t$, the right hand side term of equation \eqref{eq:A-C_discrete_fft} depends only on values of the previous time step and therefore can be solved directly in Fourier space for each frequency, leading to

\begin{align}
  \widehat{\eta}_{i}&=\frac{ \widehat{\eta}_{i}^{(t)}-\dfrac{L}{V_{m}} \Delta t \widehat{\dfrac{\partial f_c^{(t)}}{\partial \eta_{i}}}}{\bigg{(} 1+\dfrac{L}{V_{m}} \Delta t \kappa_{\eta} \|\boldsymbol{\xi}\|^2 \bigg{)} }
  \label{eq:A-C_discrete_fft_1}
\end{align}


\subsection{Preconditioning and resolution}
\label{appendix:resol}

The discretised Eq.~\eqref{eq:C-H_discrete_fft} corresponds to a linear  algebraical 
equations $\widehat{\mathcal{A}}_{c_{(j)}}(\widehat{\delta c})=\widehat{b}_{c_{(j)}}$ in which the linear operator is

\begin{align}
  \widehat{\mathcal{A}}_{c_{(j)}}(\cdot) =& (\cdot) -V_{m}\Delta t \ i \boldsymbol{\xi} \cdot \mathcal{F}\bigg{\{} M_{c_{(j)}} \mathcal{F}^{-1} \bigg{[} i \boldsymbol{\xi} \bigg{[}\mathcal{F}\bigg{(}\dfrac{\partial^{2} f_{c_{(j)}}}{\partial c^{2}}\mathcal{F}^{-1}(\cdot)\bigg{)}+\kappa_c \|\boldsymbol{\xi}\|^{2} (\cdot) \bigg{]} \bigg{]}+ \nonumber
  \\
  &\dfrac{d M_{c_{(j)}}}{d c} \mathcal{F}^{-1} (\cdot) \mathcal{F}^{-1} \bigg{[}i \boldsymbol{\xi} \bigg{(}\widehat{\dfrac{\partial f_{c_{(j)}}}{\partial c}} + \kappa_c \|\boldsymbol{\xi}\|^{2} \widehat{c}_{(j)} \bigg{)} \bigg{]} \bigg{\}}
\end{align}

\noindent and the right-hand-side (independent term) corresponds to

\begin{align}
  &\widehat{b}_{c_{(j)}} = (\widehat{c}^{(t)}-\widehat{c}_{(j)}) +V_{m}\Delta t \ i \boldsymbol{\xi} \cdot \mathcal{F} \bigg{\{} M_{c_{(j)}} \mathcal{F}^{-1} \bigg{[} i\boldsymbol{\xi} \bigg{(}\widehat{\dfrac{\partial f_{c_{(j)}}}{\partial c}}+\kappa_c \|\boldsymbol{\xi}\|^{2} \widehat{c}_{(j)} \bigg{)} \bigg{]} \bigg{\}}
  \label{eq:C-H_fft_b}
\end{align}

The linear system defined by these equations for each time step can be solved by an iterative Krylov method, e.g. the Conjugate Gradient method, thus avoiding to form and store the matrix representing the linear operator. To accelerate the convergence of the iterative solver, an approximate inverse operator is used as preconditioner in the preconditioned conjugate gradient method (PCG)  \cite{barrett_1994}. Following the procedure of Ref.~\cite{LUCARINI2019103131}, the preconditioner operator $\widehat{\mathcal{M}}_{c_{(j)}}$ is defined as:

\begin{align}
  \widehat{\mathcal{M}}_{c_{(j)}}(\cdot) = (\cdot) \bigg{[} 1+V_{m}\Delta t \overline{M}_{c_{j}} \bigg{(}\overline{\dfrac{\partial^{2} f_{c_{(j)}}}{\partial c^{2}}} \|\boldsymbol{\xi}\|^{2} +\kappa_c \|\boldsymbol{\xi}\|^{4} \bigg{)} \bigg{]}
  \label{eq:precondition}
\end{align}
where $\overline{M}_{c_{(j)}}$ and $\overline{\partial^{2} f_{c_{(j)}}/\partial c^{2}}$ are the average values of $M_{c_{(j)}}$ and $\partial^{2} f_{c_{(j)}}/\partial c^{2}$
over the domain.

For every time step, the iterations are performed until $\widehat{c}_{(j+1)}$ is converged. Since $M_{c_{(j+1)}}$ and $\partial f_{c_{(j+1)}}/\partial c$ were linearised in Eq.~\eqref{eq:C-H_discrete_1}, Eq.~\eqref{eq:C-H_discrete} is only satisfied for $c_{(j+1)}$ after convergence. The difference between left and right hand sides of Eq.~\eqref{eq:C-H_discrete} is used as a residual,

\begin{align}
  \text{Res}_{(j)} = \mathcal{F}^{-1} \bigg{\{}\dfrac{\widehat{c}_{(j+1)}-\widehat{c}^{(t)}}{V_{m}^{2}\Delta t}-i \boldsymbol{\xi} \cdot \mathcal{F}\bigg{\{} \dfrac{M_{c_{(j+1)}}}{V_{m}}\mathcal{F}^{-1}\bigg{[}i\boldsymbol{\xi} \bigg{(}\widehat{\dfrac{\partial f_{c_{(j+1)}}}{\partial c}}+\kappa_c\|\boldsymbol{\xi}\|^{2} \widehat{c}_{(j+1)}\bigg{)}\bigg{]}\bigg{\}} \bigg{\}}.
\end{align}
The absolute error for the current value of  $\widehat{c}_{(j+1)}$ is then given by the norm of the residual $\text{Err}_{(j)}=\int_{V} \text{Res}_{(j)}^{2}dV$ and the equation converges when this error becomes smaller than the tolerance. 

Variable time stepping is performed in the following way. When a large number of iterations
is reached in the Newton method ($j_{\rm Newton}>it_{\rm max}$, with $it_{\rm max}=200$) and the solution is not
converged, $\Delta t$ is reduced by a factor of 2 in order to stabilise the algorithm. The time step
is also reduced by a factor of 2 when the PCG exceeds a high number of iterations ($j_\text{PCG}>it_{\rm max}$).
When the number of iterations is below $it_{\rm max}$, the new value of time step is computed as
$\Delta t^{(t+\Delta t)}=\tau \Delta t^{(t)}$, where $\tau$ is a time step factor computed as $\tau=\min(\tau_{\rm algorithm})$,
with $\tau_{\rm algorithm}=(it_{\rm target}/it_{\rm algorithm})^{0.3}$, $it_{\rm target}=25$ the target number of iterations to get
convergence, $it_{\rm algorithm=Newton}$ the number of iterations needed to get the solution in the previous time step by
the Newton method, and $it_{\rm algorithm=PCG}$ the maximum number of iterations needed by the PCG method to converge in the previous time step.

\newpage
\subsection{Algorithms}
\label{appendix:algo}

The resolution scheme of the complete system of equations is presented in Algorithm~\ref{alg:fft-based_solution} and the
resolution of the concentration field in Algorithm~\ref{alg:c_solution}.

\RestyleAlgo{ruled}
\SetKwComment{Comment}{/* }{ */}
\begin{algorithm}[h!]
  \caption{Phase-Field resolution algorithm for 2D domain}
  Fields initialisation\;
  Time discretisation $[0,r \Delta t,...,t_{\rm max}]$, for $r \in \mathbb{Z}$\;
  Space discretisation $x \leftarrow [0,s \Delta x,...,l_{x}]$, $y \leftarrow [0,s \Delta y,...,l_{y}]$, for $s \in \mathbb{Z}$\;
  $\boldsymbol{\xi}$ and $\|\boldsymbol{\xi}\|^2$ with Eqs.~\eqref{eq:grad_fft-fdO4} and \eqref{eq:Gamma_fft-fdO4}, respectively\;
  \While{$t < t_{\rm max}$}{
    Concentration field computation:\\
    $c^{(t+\Delta t)}$ with Algorithm \ref{alg:c_solution}\;
    Phase field computation:\\
    $\dfrac{\partial f_c^{(t)}}{\partial \eta_{i}}$\;
    $\widehat{\dfrac{\partial f_c^{(t)}}{\partial \eta_{i}}} \xleftarrow{\mathcal{F}} \dfrac{\partial f_c^{(t)}}{\partial \eta_{i}}$\;     
    $\widehat{\eta}_{i}^{(t+\Delta t)} \leftarrow \frac{\widehat{\eta}_{i}^{(t)}-\dfrac{L}{V_{m}} \Delta t \widehat{\dfrac{\partial f_c^{(t)}}{\partial \eta_{i}}}}{\bigg{(}1+\dfrac{L}{V_{m}}\Delta t \kappa_{\eta} \|\boldsymbol{\xi}\|^2 \bigg{)}}$\;
    ${\eta_{i}}^{(t+\Delta t)} \xleftarrow{\mathcal{F}^{-1}} \widehat{\eta_{i}}^{(t+\Delta t)}$\;
  }
  \textbf{Output generation}\;
  \label{alg:fft-based_solution}
\end{algorithm}

\newpage

\RestyleAlgo{ruled}
\SetKwComment{Comment}{/* }{ */}
\begin{algorithm}[h!]
  \caption{$c$ resolution algorithm for 2D domain}
  \textbf{Variable initialisations:}\\
  $\text{Err}_{tol} \xleftarrow{} 10^{-10}$\;
  $\widehat{c}_{(j=1)}^{(t+\Delta t)} \xleftarrow{\mathcal{F}} c^{(t)}$\;
  \While{${\rm Err}_{(j)} > {\rm Err}_{tol}$}{
    $M_{c_{(j)}}^{(t+\Delta t)}; \ \dfrac{dM_{c_{(j)}}^{(t+\Delta t)}}{d c}$\;
    $\dfrac{\partial f_{c_{(j)}}^{(t+\Delta t)}}{\partial c}; \ \dfrac{\partial^{2} f_{c_{(j)}}^{(t+\Delta t)}}{\partial c^{2}}$\;
    $\widehat{\dfrac{\partial f_{c_{(j)}}^{(t+\Delta t)}}{\partial c}} \xleftarrow{\mathcal{F}} \dfrac{\partial f_{c_{(j)}}^{(t+\Delta t)}}{\partial c}$\;
    $\widehat{b}_{c_{(j)}}^{(t)} \xleftarrow{} \big{(}\widehat{c}^{(t)}-\widehat{c}_{(j)}^{(t+\Delta t)}\big{)} +V_{m}\Delta t \ i \boldsymbol{\xi} \cdot \mathcal{F} \bigg{\{} M_{c_{(j)}}^{(t+\Delta t)} \mathcal{F}^{-1} \bigg{[} i\boldsymbol{\xi} \bigg{(}\widehat{\dfrac{\partial f_{c_{(j)}}^{(t+\Delta t)}}{\partial c}}+\kappa_c \|\boldsymbol{\xi}\|^{2} \widehat{c}_{(j)}^{(t+\Delta t)} \bigg{)} \bigg{]} \bigg{\}}$\;
    $\widehat{\Delta c}_{(j)}^{(t+\Delta t)} \xleftarrow{} \widehat{\mathcal{A}}_{c_{(j)}}^{-1}(\widehat{b}_{c_{(j)}})$ computed with PCG method and $\widehat{\mathcal{M}}_{c_{(j)}}$ with Eq. \eqref{eq:precondition}\;
    $\widehat{c}_{(j+1)}^{(t+\Delta t)} \xleftarrow{} \widehat{c}_{(j)}^{(t+\Delta t)}+\widehat{\Delta c}_{(j)}^{(t+\Delta t)}$\;
    $c_{(j+1)}^{(t+\Delta t)} \xleftarrow{\mathcal{F}^{-1}} \widehat{c}_{(j+1)}^{(t+\Delta t)}$\;
    $\text{Res}_{(j)} \xleftarrow{\mathcal{F}^{-1}} \dfrac{\big{(}\widehat{c}_{(j+1)}^{(t+\Delta t)}-\widehat{c}^{(t)}\big{)}}{V_{m}^{2}\Delta t}-i \boldsymbol{\xi} \cdot \mathcal{F}\bigg{\{} \dfrac{M_{c_{(j+1)}}^{(t+\Delta t)}}{V_{m}}\mathcal{F}^{-1}\bigg{[}i\boldsymbol{\xi} \bigg{(}\widehat{\dfrac{\partial f_{c_{(j+1)}}^{(t+\Delta t)}}{\partial c}}+\kappa_c\|\boldsymbol{\xi}\|^{2} \widehat{c}_{(j+1)}^{(t+\Delta t)}\bigg{)}\bigg{]}\bigg{\}}$\;
    $\text{Err}_{(j)} \xleftarrow{} \int_{V} \text{Res}_{(j)}^{2} dV$\;
    $j \leftarrow j+1$\;
  }
  $c^{(t+\Delta t)} \xleftarrow{} c_{(j)}^{(t+\Delta t)}$\;
  \label{alg:c_solution}
\end{algorithm}

\newpage

\bibliographystyle{ieeetr}
\bibliography{ref}

\begin{thebibliography}{100}

\bibitem{bkm:RefLeyens2003-1}
C.~Leyens and M.~Peters, {\em Titanium and titanium alloys: Fundamentals and
  applications}.
\newblock John Wiley \& Sons, 2003.

\bibitem{bkm:RefBanerjee2013-2}
D.~Banerjee and J.~Williams, ``Perspectives on titanium science and
  technology,'' {\em Acta Materialia}, vol.~61, p.~844, 2013.

\bibitem{bkm:RefFrasier2014-3}
W.~Frazier, ``Metal additive manufacturing: A review,'' {\em Journal of
  Materials Engineering and Performance}, vol.~23, pp.~1917--1928, 2014.

\bibitem{bkm:refHerzog2016-4}
D.~Herzog, V.~Seyda, E.~Wycisk, and C.~Emmelmann, ``Additive manufacturing of
  metals,'' {\em Acta Materialia}, vol.~117, pp.~371--392, 2016.

\bibitem{bkm:RefBecker2021-5}
T.~Becker, P.~Kumar, and U.~Ramamurty, ``Fracture and fatigue in additively
  manufactured metals,'' {\em Acta Materialia}, vol.~219, p.~117240, 2021.

\bibitem{bkm:RefVandenbrouke2007-6}
B.~Vandenbroucke and J.-P. Kruth, ``Selective laser melting of biocompatible
  metals for rapid manufacturing of medical parts,'' {\em Rapid Prototyping
  Journal}, vol.~13, pp.~196--203, 2007.

\bibitem{bkm:RefFacchini2010-7}
L.~Facchini, E.~Magalini, P.~Robotti, A.~Molinari, S.~H\"oges, and
  K.~Wissenbach, ``Ductility of a \uppercase{T}i-6\uppercase{A}l-4\uppercase{V}
  alloy produced by selective laser melting of prealloyed powders,'' {\em Rapid
  Prototyping Journal}, vol.~16, pp.~450--459, 2010.

\bibitem{bkm:RefVrancken2012-8}
B.~Vrancken, L.~Thijs, J.-P. Kruth, and J.~Humbeeck, ``Heat treatment of
  \uppercase{T}i6\uppercase{A}l4\uppercase{V} produced by selective laser
  melting: Microstructure and mechanical properties,'' {\em Journal of Alloys
  and Compounds}, vol.~541, pp.~177--185, 2012.

\bibitem{bkm:RefMurr2009-9}
L.~Murr, S.~Quinones, S.~Gaytan, M.~Lopez, A.~Rodela, E.~Martinez,
  D.~Hernandez, E.~Martinez, F.~Medina, and R.~Vicker, ``Microstructure and
  mechanical behavior of \uppercase{T}i-6\uppercase{}al-4\uppercase{V} produced
  by rapid-layer manufacturing, for biomedical applications,'' {\em Journal of
  the Mechanical Behavior of Biomedical Materials}, vol.~2, pp.~20--32, 2009.

\bibitem{bkm:RefVilaro2011-10}
T.~Vilaro, C.~Colin, and J.~Bartout, ``As-fabricated and heat-treated
  microstructures of the \uppercase{T}i-6\uppercase{A}l-4\uppercase{V} alloy
  processed by selective laser melting,'' vol.~42, pp.~3190--3199, 2011.

\bibitem{bkm:RefXu2015-11}
W.~Xu, M.~Brandt, S.~Sun, J.~Elambasseril, Q.~Liu, K.~Latham, K.~Xia, and
  M.Qian, ``Additive manufacturing of strong and ductile
  \uppercase{T}i-6\uppercase{A}l-4\uppercase{V} by selective laser melting via
  in situ martensite decomposition,'' {\em Acta Materialia}, vol.~85,
  pp.~74--84, 2015.

\bibitem{bkm:RefXu2016-12}
W.~Xu, E.~Lui, A.Pateras, M.~Qian, and M.Brandt, ``In situ tailoring
  microstructure in additively manufactured
  \uppercase{T}i-6\uppercase{A}l-4\uppercase{V} for superior mechanical
  performance,'' {\em Acta Materialia}, vol.~125, pp.~390--400, 2017.

\bibitem{bkm:RefTerHaar2018-13}
G.~T. Haar and T.~Becker, ``Selective laser melting produced
  \uppercase{T}i-6\uppercase{A}l-4\uppercase{V}: Post-process heat treatments
  to achieve superior tensile properties,'' {\em Materials}, vol.~11, p.~146,
  2018.

\bibitem{bkm:RefKaschel2020a-14}
F.~Kaschel, M.~Celikin, and D.~Dowling, ``Effects of laser power on geometry,
  microstructure and mechanical properties of printed
  \uppercase{T}i-6\uppercase{A}l-4\uppercase{V} parts,'' {\em Journal of
  Materials Processing Technology}, vol.~278, p.~116539, 2020.

\bibitem{bkm:RefWelsch1998-15}
G.~Welsch, Boyer, and E.~Collings, {\em Materials properties handbook: Titanium
  alloys}.
\newblock ASM international, 2nd~ed., 1998.

\bibitem{bkm:RefElmer2005-16}
J.~Elmer, T.~Palmer, S.~Babu, and E.~Specht, ``In situ observations of lattice
  expansion and transformation rates of $\alpha$ and $\beta$ phases in
  \uppercase{T}i-6\uppercase{A}l-4\uppercase{V},'' {\em Materials Science and
  Engineering: A}, vol.~391, pp.~104--113, 2005.

\bibitem{bkm:RefKolichev1999-17}
B.~Kolichev, V.I.Elagin, and V.~Livanov, {\em Metallurgy and heat treatment of
  non-ferrous metals and alloys}.
\newblock Moscow, Russia: MISIS, 1999.

\bibitem{bkm:RefWilliams1970-18}
J.~Williams and B.~Hickman, ``Tempering behavior of orthorhombic martensite in
  titanium alloys,'' {\em Metallurgical Transactions}, vol.~1, pp.~2648--2650,
  1970.

\bibitem{bkm:RefFroes2015-19}
F.~Froes, {\em Titanium: Physical metallurgy, processing, and applications}.
\newblock Ohio: ASM International, Materials Park, 2015.

\bibitem{bkm:RefBoyer1994-20}
R.~Boyer, G.~Welsch, and E.~Collings, {\em Materials properties handbook:
  Titanium alloys}.
\newblock ASM International, 1994.

\bibitem{bkm:RefZafari2018-21}
A.~Zafari and K.~Xia, ``High ductility in a fully martensitic microstructure: A
  paradox in a \uppercase{T}i alloy produced by selective laser melting,'' {\em
  Materials Research Letters}, vol.~6, pp.~627--633, 2018.

\bibitem{bkm:RefHaubrich2019-22}
J.~Haubrich, J.~Gussone, P.~Barriobero-Vila, P.~K\"urnsteiner, E.~J\"agle,
  D.~Raabe, N.~Schell, and G.~Requena, ``The role of lattice defects, element
  partitioning and intrinsic heat effects on the microstructure in selective
  laser melted \uppercase{T}i-6\uppercase{A}l-4\uppercase{V},'' {\em Acta
  Materialia}, vol.~167, pp.~136--148, 2019.

\bibitem{bkm:RefThijs2010-23}
L.~Thijs, F.~Verhaeghe, T.~Craeghs, J.~V. Humbeeck, and J.~Kruth, ``A study of
  the micro structural evolution during selective laser melting of
  \uppercase{T}i-6\uppercase{A}l-4\uppercase{V},'' {\em Acta Materialia},
  vol.~58, pp.~3303--3312, 2010.

\bibitem{bkm:RefSercombe2008-24}
T.~Sercombe, N.~Jones, R.~Day, and A.~Kop, ``Heat treatment of
  \uppercase{T}i-6\uppercase{A}l-7\uppercase{N}b components produced by
  selective laser melting,'' {\em Rapid Prototyping Journal}, vol.~14,
  pp.~300--304, 2008.

\bibitem{bkm:RefSong2012-25}
B.~Song, S.~Dong, B.~Zhang, H.~Liao, and C.~Coddet, ``Effects of processing
  parameters on microstructure and mechanical property of selective laser
  melted \uppercase{T}i6\uppercase{A}l4\uppercase{V},'' {\em Materials \&
  Design}, vol.~35, pp.~120--125, 2012.

\bibitem{bkm:RefWielewski2012-26}
E.~Wielewski, C.~Siviour, and N.~Petrinic, ``On the correlation between
  macrozones and twinning in \uppercase{T}i-6\uppercase{A}l-4\uppercase{V} at
  very high strain rates,'' {\em Scripta Materialia}, vol.~67, pp.~229--232,
  2012.

\bibitem{bkm:RefSimonelli2014-27}
M.~Simonelli, Y.~Tse, and C.~Tuck, ``The formation of $\alpha+\beta$
  microstructure in as-fabricated selective laser melting of
  \uppercase{T}i-6\uppercase{A}l-4\uppercase{V},'' {\em Journal of Materials
  Research}, vol.~29, pp.~2028--2035, 2014.

\bibitem{bkm:RefYang2016-28}
J.~Yang, H.~Yu, J.~Yin, M.~Gao, Z.~Wang, and X.~Zeng, ``Formation and control
  of martensite in \uppercase{T}i-6\uppercase{A}l-4\uppercase{V} alloy produced
  by selective laser melting,'' {\em Materials \& Design}, vol.~108,
  pp.~308--318, 2016.

\bibitem{bkm:RefWu2016-29}
S.~Wu, Y.~Lu, Y.~Gan, T.~Huang, C.~Zhao, J.~Lin, S.~Guo, and J.~Lin,
  ``Microstructural evolution and microhardness of a selective-laser-melted
  \uppercase{T}i-6\uppercase{A}l-4\uppercase{V} alloy after post heat
  treatments,'' {\em Journal of Alloys and Compounds}, vol.~672, pp.~643--652,
  2016.

\bibitem{bkm:RefKasperovich2015-30}
G.~Kasperovich and J.~Hausmann, ``Improvement of fatigue resistance and
  ductility of \uppercase{T}i\uppercase{A}l6\uppercase{V}4 processed by
  selective laser melting,'' {\em Journal of Materials Processing Technology},
  vol.~220, pp.~202--214, 2015.

\bibitem{bkm:RefTan2016-31}
X.~Tan, Y.~Kok, W.~Toh, Y.~Tan, M.~Descoins, D.~Mangelinck, S.~Tor, K.~Leong,
  and C.~Chua, ``Revealing martensitic transformation and alpha/beta interface
  evolution in electron beam melting three-dimensional-printed
  \uppercase{T}i-6\uppercase{A}l-4\uppercase{V},'' {\em Scientific Reports},
  vol.~6, p.~26039, 2016.

\bibitem{bkm:RefKrakhmalev2016-32}
P.~Krakhmalev, G.~Fredriksson, I.~Yadroitsava, N.~Kazantseva, A.~du~Plessis,
  and I.~Yadroitsev, ``Deformation behavior and microstructure of
  \uppercase{T}i6\uppercase{A}l4\uppercase{V} manufactured by
  \uppercase{SLM},'' {\em Physics Procedia}, vol.~83, pp.~778--788, 2016.

\bibitem{bkm:RefHuang2016-33}
Q.~Huang, N.~Hu, X.~Yang, R.~Zhang, and Q.~Feng, ``Microstructure and inclusion
  of \uppercase{T}i-6\uppercase{A}l-4\uppercase{V} fabricated by selective
  laser melting,'' {\em Frontiers of Materials Science}, vol.~10, pp.~428--431,
  2016.

\bibitem{bkm:RefBarrioberoVila2017-34}
P.~Barriobero-Vila, J.~Gussone, J.~Haubrich, S.~Sandl\"obes, J.~D. Silva,
  P.~Cloetens, N.~Schell, and G.~Requena, ``Inducing stable $\alpha+\beta$
  microstructures during selective laser melting of
  \uppercase{T}i-6\uppercase{A}l-4\uppercase{V} using intensified intrinsic
  heat treatments,'' {\em Materials}, vol.~10, p.~268, 2017.

\bibitem{bkm:RefCao2018-35}
S.~Cao, R.~Chu, X.~Zhou, K.~Yang, Q.~Jia, C.~Lim, A.~Huang, and X.~Wu, ``Role
  of martensite decomposition in tensile properties of selective laser melted
  \uppercase{T}i-6\uppercase{A}l-4\uppercase{V},'' {\em Journal of Alloys and
  Compounds}, vol.~744, pp.~357--363, 2018.

\bibitem{bkm:RefZhang2018-36}
X.~Zhang, G.~Fang, S.~Leeflang, A.~Bottger, A.~Zadpoor, and J.~Zhou, ``Effect
  of subtransus heat treatment on the microstructure and mechanical properties
  of additively manufactured \uppercase{T}i-6\uppercase{A}l-4\uppercase{V}
  alloy,'' {\em Journal of Alloys and Compounds}, vol.~735, pp.~1562--1575,
  2018.

\bibitem{bkm:RefHumbert1995-37}
M.~Humbert, F.~Wagner, H.~Moustahfid, and C.~Esling, ``Determination of the
  orientation of a parent $\beta$ grain from the orientations of the inherited
  $\alpha$ plates in the phase transformation from body-centred cubic to
  hexagonal close packed,'' {\em Journal of Applied Crystallography}, vol.~28,
  pp.~571--576, 1995.

\bibitem{bkm:RefHumbert1996-38}
M.~Humbert, N.~Gey, J.~Muller, and C.~Esling, ``Determination of a mean
  orientation from a cloud of orientations. application to electron
  back-scattering pattern measurements,'' 1996.

\bibitem{bkm:RefGlavicic2003-39}
M.~Glavicic, P.~Kobryn, T.~Bieler, and S.~Semiatin, ``A method to determine the
  orientation of the high-temperature beta phase from measured \uppercase{EBSD}
  data for the low-temperature alpha phase in
  \uppercase{T}i-6\uppercase{A}l-4\uppercase{V},'' {\em Materials Science and
  Engineering: A}, vol.~351, pp.~258--264, 2003.

\bibitem{bkm:RefGlavicic2003b-40}
M.~Glavicic, P.~Kobryn, T.~Bieler, and S.~Semiatin, ``An automated method to
  determine the orientation of the high-temperature beta phase from measured
  \uppercase{EBSD} data for the low-temperature alpha-phase in
  \uppercase{T}i-6\uppercase{A}l-4\uppercase{V},'' {\em Materials Science and
  Engineering: A}, vol.~346, pp.~50--59, 2003.

\bibitem{bkm:RefFormanoir2016-41}
C.~de~Formanoir, M.~Suard, R.~Dendievel, G.~Martin, and S.~Godet, ``Improving
  the mechanical efficiency of electron beam melted titanium lattice structures
  by chemical etching,'' {\em Additive Manufacturing}, vol.~11, pp.~71--76,
  2016.

\bibitem{bkm:RefKarami2020-42}
K.~Karami, A.~Blok, L.~Weber, S.~Ahmadi, R.~Petrov, K.~Nikolic, E.~Borisov,
  S.~Leeflang, C.~Ayas, A.~Zadpoor, M.~Mehdipour, E.~Reinton, and V.~Popovich,
  ``Continuous and pulsed selective laser melting of \uppercase{Ti6Al4V}
  lattice structures: Effect of post-processing on microstructural anisotropy
  and fatigue behaviour,'' {\em Additive Manufacturing}, vol.~36, p.~101433,
  2020.

\bibitem{bkm:RefPantawane2021-43}
M.~Pantawane, T.~Yang, Y.~Jin, S.~Joshi, S.~Dasari, A.~Sharma, A.~Krokhin,
  S.~Srinivasan, R.~Banerjee, A.~Neogi, and N.~Dahotre, ``Crystallographic
  texture dependent bulk anisotropic elastic response of additively
  manufactured \uppercase{Ti6Al4V},'' {\em Scientific Reports}, vol.~11,
  p.~633, 2021.

\bibitem{bkm:RefDonachie2000-44}
M.~Donachie, {\em Titanium: A technical guide}.
\newblock ASM International, 2000.

\bibitem{bkm:RefKaschel2021-45}
F.~Kaschel, R.~Vijayaraghavan, P.~McNally, D.~Dowling, and M.~Celikin,
  ``In-situ \uppercase{XRD} study on the effects of stress relaxation and phase
  transformation heat treatments on mechanical and microstructural behaviour of
  additively manufactured \uppercase{T}i-6\uppercase{A}l-4\uppercase{V},'' {\em
  Materials Science and Engineering: A}, vol.~819, p.~141534, 2021.

\bibitem{liu2022effect}
Y.~Liu, H.~Xu, B.~Peng, X.~Wang, S.~Li, Q.~Wang, Z.~Li, and Y.~Wang, ``Effect
  of heating treatment on the microstructural evolution and dynamic tensile
  properties of ti-6al-4v alloy produced by selective laser melting,'' {\em
  Journal of Manufacturing Processes}, vol.~74, pp.~244--255, 2022.

\bibitem{xiao2022mechanism}
Y.~Xiao, L.~Lan, S.~Gao, B.~He, and Y.~Rong, ``Mechanism of ultrahigh ductility
  obtained by globularization of $\alpha$gb for additive manufacturing
  ti--6al--4v,'' {\em Materials Science and Engineering: A}, vol.~858,
  p.~144174, 2022.

\bibitem{dhekne2023micro}
P.~P. Dhekne, T.~Vermeij, V.~Devulapalli, S.~D. Jadhav, J.~P. Hoefnagels, M.~G.
  Geers, and K.~Vanmeensel, ``Micro-mechanical deformation behavior of
  heat-treated laser powder bed fusion processed ti-6al-4v,'' {\em Scripta
  Materialia}, vol.~233, p.~115505, 2023.

\bibitem{li2023optimizing}
Z.~Li, K.~Ming, B.~Li, S.~He, B.~Miao, and S.~Zheng, ``Optimizing
  strength-ductility of laser powder bed fusion-fabricated ti--6al--4v via
  twinning and phase transformation dominated interface engineering,'' {\em
  Materials Science and Engineering: A}, vol.~882, p.~145484, 2023.

\bibitem{bkm:RefLiu2019-58}
S.~Liu and Y.~Shin, ``Additive manufacturing of \uppercase{Ti6Al4V} alloy: A
  review,'' {\em Materials \& Design}, vol.~164, p.~107552, 2019.

\bibitem{bkm:RefHemmasian2019-59}
A.~H. Ettefagh, C.~Zeng, S.~Guo, and J.~Raush, ``Corrosion behavior of
  additively manufactured \uppercase{T}i-6\uppercase{A}l-4\uppercase{V} parts
  and the effect of post annealing,'' {\em Additive Manufacturing}, vol.~28,
  pp.~252--258, 2019.

\bibitem{bkm:RefZhang2021electrochem-60}
Y.~Zhang, L.~Feng, T.~Zhang, H.~Xu, and J.~Li, ``Heat treatment of additively
  manufactured \uppercase{T}i-6\uppercase{A}l-4\uppercase{V} alloy:
  Microstructure and electrochemical properties,'' {\em Journal of Alloys and
  Compounds}, vol.~888, p.~161602, 2021.

\bibitem{bkm:RefWang2016-61}
M.~Wang, Y.~Wu, S.~Lu, T.~Chen, Y.~Zhao, H.~Chen, and Z.~Tang, ``Fabrication
  and characterization of selective laser melting printed
  \uppercase{T}i-6\uppercase{A}l-4\uppercase{V} alloys subjected to heat
  treatment for customized implants design,'' {\em Progress in Natural Science:
  Materials International}, vol.~26, pp.~671--677, 2016.

\bibitem{bkm:RefLeuders2014-46}
S.~Leuders, T.~Lieneke, S.~Lammers, T.~Troster, and T.~Niendorf, ``On the
  fatigue properties of metals manufactured by selective laser melting-the role
  of ductility,'' {\em Journal of Materials Research}, vol.~29, pp.~1911--1919,
  2014.

\bibitem{bkm:RefKasperovich2015-47}
G.~Kasperovich and J.~Hausmann, ``Improvement of fatigue resistance and
  ductility of \uppercase{TiAl6V4} processed by selective laser melting,'' {\em
  Journal of Materials Processing Technology}, vol.~220, pp.~202--214, 2015.

\bibitem{bkm:RefGalarraga2016-48}
H.~Galarraga, D.~Lados, R.~Dehoff, M.~Kirka, and P.~Nandwana, ``Effects of the
  microstructure and porosity on properties of
  \uppercase{T}i-6\uppercase{A}l-4\uppercase{V} \uppercase{ELI} alloy
  fabricated by electron beam melting \uppercase{(EBM)},'' {\em Additive
  Manufacturing}, vol.~10, pp.~47--57, 2016.

\bibitem{bkm:RefGalarraga2017-49}
H.~Galarraga, R.~Warren, D.~Lados, R.~Dehoff, M.~Kirka, and P.~Nandwana,
  ``Effects of heat treatments on microstructure and properties of
  \uppercase{T}i-6\uppercase{A}l-4\uppercase{V} \uppercase{ELI} alloy
  fabricated by electron beam melting \uppercase{(EBM)},'' {\em Materials
  Science and Engineering: A}, vol.~685, pp.~417--428, 2017.

\bibitem{bkm:RefBaker2017-50}
A.~Baker, P.~Collins, and J.~Williams, ``New nomenclatures for heat treatments
  of additively manufactured titanium alloys,'' {\em JOM}, vol.~69,
  pp.~1221--1227, 2017.

\bibitem{bkm:RefZou2020-51}
Z.~Zou, M.~Simonelli, J.~Katrib, G.~Dimitrakis, and R.~Hague, ``Refinement of
  the grain structure of additive manufactured titanium alloys via epitaxial
  recrystallization enabled by rapid heat treatment,'' {\em Scripta
  Materialia}, vol.~180, pp.~66--70, 2020.

\bibitem{bkm:RefWycisk2014-52}
E.~Wycisk, A.~Solbach, S.~Siddique, D.~Herzog, F.~Walther, and C.~Emmelmann,
  ``Effects of defects in laser additive manufactured
  \uppercase{T}i-6\uppercase{A}l-4\uppercase{V} on fatigue properties,'' {\em
  Physics Procedia}, vol.~56, pp.~371--378, 2014.

\bibitem{bkm:RefZeng2005-62}
L.~Zeng and T.~Bieler, ``Effects of working, heat treatment, and aging on
  microstructural evolution and crystallographic texture of $\alpha$, $\alpha
  {}'$, $\alpha {}' {}'$ and $\beta$ phases in
  \uppercase{T}i-6\uppercase{A}l-4\uppercase{V} wire,'' {\em Materials Science
  and Engineering: A}, vol.~392, pp.~403--414, 2005.

\bibitem{bkm:RefAlBermani2010-63}
S.~Al-Bermani, M.~Blackmore, W.~Zhang, and I.~Todd, ``The origin of
  microstructural diversity, texture, and mechanical properties in electron
  beam melted \uppercase{T}i-6\uppercase{A}l-4\uppercase{V},'' {\em
  Metallurgical and Materials Transactions A}, vol.~41, pp.~3422--3434, 2010.

\bibitem{bkm:RefSallicaLeva2016-64}
E.~Sallica-Leva, R.~Caram, A.~Jardini, and J.~Fogagnolo, ``Ductility
  improvement due to martensite $\alpha {}'$ decomposition in porous
  \uppercase{T}i-6\uppercase{A}l-4\uppercase{V} parts produced by selective
  laser melting for orthopedic implants,'' {\em Journal of the Mechanical
  Behavior of Biomedical Materials}, vol.~54, pp.~149--158, 2016.

\bibitem{bkm:RefKazantseva2018-65}
N.~Kazantseva, P.~Krakhmalev, M.~Thuvander, I.~Yadroitsev, N.~Vinogradova, and
  I.~Ezhov, ``Martensitic transformations in
  \uppercase{T}i-6\uppercase{A}l-4\uppercase{V} (\uppercase{ELI}) alloy
  manufactured by 3\uppercase{D} printing,'' {\em Materials Characterization},
  vol.~146, pp.~101--112, 2018.

\bibitem{bkm:RefGupta2016-66}
R.~Gupta, V.~A. Kumar, and S.~Chhangani, ``Study on variants of solution
  treatment and aging cycle of titanium alloy \uppercase{Ti6Al4V},'' {\em
  Journal of Materials Engineering and Performance}, vol.~25, pp.~1492--1501,
  2016.

\bibitem{bkm:RefKaschel2020b-67}
F.~Kaschel, R.~Vijayaraghavan, A.~Shmeliov, E.~McCarthy, M.~Canavan,
  P.~McNally, D.~Dowling, V.~Nicolosi, and M.~Celikin, ``Mechanism of stress
  relaxation and phase transformation in additively manufactured
  \uppercase{T}i-6\uppercase{A}l-4\uppercase{V} via in situ high temperature
  \uppercase{XRD} and \uppercase{TEM} analyses,'' {\em Acta Materialia},
  vol.~188, pp.~720--732, 2020.

\bibitem{bkm:RefChao2014-68}
Q.~Chao, P.~Hodgson, and H.~Beladi, ``Ultra-fine grain formation in a
  \uppercase{T}i-6\uppercase{A}l-4\uppercase{V} alloy by thermomechanical
  processing of a martensitic microstructure,'' {\em Metallurgical and
  Materials Transactions A}, vol.~45, pp.~2659--2671, 2014.

\bibitem{bkm:RefGhosh2022-69}
A.~Ghosh, V.~Sahu, and N.~Gurao, ``Effect of heat treatment on the ratcheting
  behaviour of additively manufactured and thermo-mechanically treated
  \uppercase{T}i-6\uppercase{A}l-4\uppercase{V} alloy,'' {\em Materials Science
  and Engineering: A}, vol.~833, p.~142345, 2022.

\bibitem{bkm:RefSabban2019-70}
R.~Sabban, S.~Bahl, K.~Chatterjee, and S.~Suwas, ``Globularization using heat
  treatment in additively manufactured
  \uppercase{T}i-6\uppercase{A}l-4\uppercase{V} for high strength and
  toughness,'' {\em Acta Materialia}, vol.~162, pp.~239--254, 2019.

\bibitem{bkm:RefChen2011-71}
J.~Chen and W.~Tsai, ``In situ corrosion monitoring of
  \uppercase{T}i-6\uppercase{A}l-4\uppercase{V} alloy in \uppercase{H2SO4/HCl}
  mixed solution using electrochemical \uppercase{AFM},'' {\em Electrochimica
  Acta}, vol.~56, no.~4, pp.~1746--1751, 2011.

\bibitem{bkm:RefCalta2020-72}
N.~Calta, V.~Thampy, D.~Lee, A.~Martin, R.~Ganeriwala, J.~Wang, P.~Depond,
  T.~Roehling, A.~Fong, A.~Kiss, C.~Tassone, K.~Stone, J.~N. Weker, M.~Toney,
  A.~V. Buuren, and M.~Matthews, ``Cooling dynamics of two titanium alloys
  during laser powder bed fusion probed with in situ \uppercase{X}-ray imaging
  and diffraction,'' {\em Materials \& Design}, vol.~195, p.~108987, 2020.

\bibitem{bkm:RefGilMur1996-73}
F.~G. Mur, D.~Rodr\'iguez, and J.~Planell, ``Influence of tempering temperature
  and time on the $\alpha {}'$ \uppercase{T}i-6\uppercase{A}l-4\uppercase{V}
  martensite,'' {\em Journal of Alloys and Compounds}, vol.~234, pp.~287--289,
  1996.

\bibitem{bkm:RefMurgau2012-74}
C.~C. Murgau, R.~Pederson, and L.~Lindgren, ``A model for
  \uppercase{T}i-6\uppercase{A}l-4\uppercase{V} microstructure evolution for
  arbitrary temperature changes,'' {\em Modelling and Simulation in Materials
  Science and Engineering}, vol.~20, p.~055006, 2012.

\bibitem{bkm:RefYang2021-75}
X.~Yang, R.~Barrett, M.~Tong, N.~Harrison, and S.~Leen, ``Towards a
  process-structure model for \uppercase{T}i-6\uppercase{A}l-4\uppercase{V}
  during additive manufacturing,'' {\em Journal of Manufacturing Processes},
  vol.~61, pp.~428--439, 2021.

\bibitem{bkm:RefBAYKASOGLU}
C.~Baykasoğlu, O.~Akyildiz, M.~Tunay, and A.~C. To, ``A process-microstructure
  finite element simulation framework for predicting phase transformations and
  microhardness for directed energy deposition of ti6al4v,'' {\em Additive
  Manufacturing}, vol.~35, p.~101252, 2020.

\bibitem{bkm:met8080633}
E.~Salsi, M.~Chiumenti, and M.~Cervera, ``Modeling of microstructure evolution
  of ti6al4v for additive manufacturing,'' {\em Metals}, vol.~8, no.~8, 2018.

\bibitem{bkm:SUN}
W.~Sun, F.~Shan, N.~Zong, H.~Dong, and T.~Jing, ``A simulation and experiment
  study on phase transformations of ti-6al-4v in wire laser additive
  manufacturing,'' {\em Materials \& Design}, vol.~207, p.~109843, 2021.

\bibitem{chen2002phase}
L.-Q. Chen, ``Phase-field models for microstructure evolution,'' {\em Annual
  review of materials research}, vol.~32, no.~1, pp.~113--140, 2002.

\bibitem{bkm:RefShi2016-77}
R.~Shi, D.~Wang, and Y.~Wang, {\em Chapter: Modeling and simulation of
  microstructure evolution during heat treatment of titanium alloys}.
\newblock ASM International, 2016.

\bibitem{bkm:RefJi2018-78}
Y.~Ji, L.~Chen, and L.-Q. Chen, {\em Chapter 6- Understanding microstructure
  evolution during additive manufacturing of metallic alloys using phase-field
  modeling}.
\newblock Butterworth-Heinemann, 2018.

\bibitem{bkm:RefTourret2022-76}
D.~Tourret, H.~Liu, and J.~LLorca, ``Phase-field modeling of microstructure
  evolution: Recent applications, perspectives and challenges,'' {\em Progress
  in Materials Science}, vol.~123, p.~100810, 2022.

\bibitem{shi2019integrated}
R.~Shi, S.~Khairallah, T.~W. Heo, M.~Rolchigo, J.~T. McKeown, and M.~J.
  Matthews, ``Integrated simulation framework for additively manufactured
  ti-6al-4v: melt pool dynamics, microstructure, solid-state phase
  transformation, and microelastic response,'' {\em Jom}, vol.~71,
  pp.~3640--3655, 2019.

\bibitem{bkm:RefHuang2019-79}
S.~Huang, J.~Zhang, Y.~Ma, S.~Zhang, S.~Youssef, M.~Qi, H.~Wang, J.~Qiu, D.~Xu,
  J.~Lei, and R.~Yang, ``Influence of thermal treatment on element partitioning
  in $\alpha+\beta$ titanium alloy,'' {\em Journal of Alloys and Compounds},
  vol.~791, pp.~575--585, 2019.

\bibitem{bkm:RefAhluwalia2020-80}
R.~Ahluwalia, R.~Laskowski, N.~Ng, M.~Wong, S.~Quek, and D.~Wu, ``Phase field
  simulation of $\alpha$/$\beta$ microstructure in titanium alloy welds,'' {\em
  Materials Research Express}, vol.~7, p.~046517, 2020.

\bibitem{bkm:RefZhang2021-81}
J.~Zhang, H.~Ju, H.~Xu, L.~Yang, Z.~Meng, C.~Liu, P.~Sun, J.~Qiu, C.~Bai,
  D.~Xu, and R.~Yang, ``Effects of heating rate on the alloy element
  partitioning and mechanical properties in equiaxed $\alpha+\beta$
  \uppercase{T}i-6\uppercase{A}l-4\uppercase{V} alloy,'' {\em Journal of
  Materials Science and Technology}, vol.~94, pp.~1--9, 2021.

\bibitem{bkm:RefBoccardo2023-91}
A.~Boccardo, M.~Tong, S.~Leen, D.~Tourret, and J.~Segurado, ``Efficiency and
  accuracy of \uppercase{GPU}-parallelized fourier spectral methods for solving
  phase-field models,'' {\em Computational Materials Science}, vol.~228,
  p.~112313, 2023.

\bibitem{bkm:RefZou2021-82}
Z.~Zou, M.~Simonelli, J.~Katrib, G.~Dimitrakis, and R.~Hague, ``Microstructure
  and tensile properties of additive manufactured
  \uppercase{T}i-6\uppercase{A}l-4\uppercase{V} with refined prior-$\beta$
  grain structure obtained by rapid heat treatment,'' {\em Materials Science
  and Engineering: A}, vol.~814, p.~141271, 2021.

\bibitem{bkm:RefMiyazaki2019-83}
S.~Miyazaki, M.~Kusano, D.~Bulgarevich, S.~Kishimoto, A.~Yumoto, and
  M.~Watanabe, ``Image segmentation and analysis for microstructure and
  property evaluations on \uppercase{T}i-6\uppercase{A}l-4\uppercase{V}
  fabricated by selective laser melting,'' {\em Materials Transactions},
  vol.~60, pp.~561--568, 2019.

\bibitem{bkm:RefSteinbach1996-84}
I.~Steinbach, F.~Pezzolla, B.~Nestler, M.~See{\ss}elberg, R.~Prieler,
  G.~Schmitz, and J.~Rezende, ``A phase field concept for multiphase systems,''
  {\em Physica D: Nonlinear Phenomena}, vol.~94, pp.~135--147, 1996.

\bibitem{bkm:RefEiken2006-85}
J.~Eiken, B.~B\"ottger, and I.~Steinbach, ``Multiphase-field approach for
  multicomponent alloys with extrapolation scheme for numerical application,''
  {\em Physical Review E}, vol.~73, p.~066122, 2006.

\bibitem{bkm:RefSteinbach2006-86}
I.~Steinbach and M.~Apel, ``Multi phase field model for solid state
  transformation with elastic strain,'' {\em Physica D: Nonlinear Phenomena},
  vol.~217, pp.~153--160, 2006.

\bibitem{bkm:RefOforiOpoku2010-87}
N.~Ofori-Opoku and N.~Provatas, ``A quantitative multi-phase field model of
  polycrystalline alloy solidification,'' {\em Acta Materialia}, vol.~58,
  pp.~2155--2164, 2010.

\bibitem{bkm:RefShi2015-77_1}
R.~Shi, N.~Zhou, S.~Niezgoda, and Y.~Wang, ``Microstructure and transformation
  texture evolution during $\alpha$ precipitation in polycrystalline $\alpha /
  \beta$ titanium alloys - \uppercase{A} simulation study,'' {\em Acta
  Materialia}, vol.~94, pp.~224--243, 2015.

\bibitem{bkm:RefLoginova2003-88}
I.~Loginova, J.~Odqvist, G.~Amberg, and J.~{\AA}gren, ``The phase-field
  approach and solute drag modeling of the transition to massive $\gamma
  \rightarrow \alpha$ transformation in binary \uppercase{Fe-C} alloys,'' {\em
  Acta Materialia}, vol.~51, pp.~1327--1339, 2003.

\bibitem{bkm:RefZhu2004-89}
J.~Zhu, T.~Wang, A.~Ardell, S.~Zhou, Z.~Liu, and L.-Q. Chen,
  ``Three-dimensional phase-field simulations of coarsening kinetics of $\gamma
  {}'$ particles in binary \uppercase{N}i-\uppercase{A}l alloys,'' {\em Acta
  Materialia}, vol.~52, pp.~2837--2845, 2004.

\bibitem{bkm:RefAnsara1998-90}
I.~Ansara, A.~Dinsdale, and M.~Rand, {\em \uppercase{COST} 507, Definition of
  thermochemical and thermophysical properties to provide a database for the
  development of new light alloys: Thermochemical database for light metal
  alloys}, vol.~2.
\newblock Publications Office, 1998.

\bibitem{barrett_1994}
R.~Barrett, M.~Berry, T.~F. Chan, J.~Demmel, J.~Donato, J.~Dongarra,
  V.~Eijkhout, R.~Pozo, C.~Romine, and H.~van~der Vorst, {\em Templates for the
  Solution of Linear Systems: Building Blocks for Iterative Methods}.
\newblock Society for Industrial and Applied Mathematics, 1994.

\bibitem{LUCARINI2019103131}
S.~Lucarini and J.~Segurado, ``Dbfft: A displacement based fft approach for
  non-linear homogenization of the mechanical behavior,'' {\em International
  Journal of Engineering Science}, vol.~144, p.~103131, 2019.

\bibitem{sk_cuda_2021}
L.~{Givon}, ``{Scikit-CUDA Documentation},'' {\em
  https://scikit-cuda.readthedocs.io/en/latest}, 2021.

\bibitem{kloeckner_pycuda_2012}
A.~{Kl{\"o}ckner}, N.~{Pinto}, Y.~{Lee}, B.~{Catanzaro}, P.~{Ivanov}, and
  A.~{Fasih}, ``{PyCUDA and PyOpenCL: A Scripting-Based Approach to GPU
  Run-Time Code Generation},'' {\em Parallel Computing}, vol.~38, no.~3,
  pp.~157--174, 2012.

\bibitem{bkm:RefSingman1984-92}
C.~Singman, ``Atomic volume and allotropy of the elements,'' {\em Journal of
  Chemical Education}, vol.~61, pp.~137--142, 1984.

\bibitem{bkm:RefGierlotka2019-93}
W.~Gierlotka, G.~Lothongkum, B.~Lohwongwatana, and C.~Puncreoburt, ``Atomic
  mobility in titanium grade 5 (\uppercase{T}i6\uppercase{A}l4\uppercase{V}),''
  {\em Journal of Mining and Metallurgy, Section B: Metallurgy}, vol.~55,
  pp.~65--77, 2019.

\bibitem{smith_2009}
M.~Smith, {\em ABAQUS/Standard User's Manual, Version 6.9}.
\newblock United States: Dassault Syst{\`e}mes Simulia Corp, 2009.

\bibitem{boccardo_2017}
A.~Boccardo, P.~Dardati, D.~Celentano, and L.~Godoy, ``Austempering heat
  treatment of ductile iron: Computational simulation and experimental
  validation,'' {\em Finite Elements in Analysis and Design}, vol.~134,
  pp.~82--91, 2017.

\bibitem{zou2021microstructure}
Z.~Zou, M.~Simonelli, J.~Katrib, G.~Dimitrakis, and R.~Hague, ``Microstructure
  and tensile properties of additive manufactured ti-6al-4v with refined
  prior-$\beta$ grain structure obtained by rapid heat treatment,'' {\em
  Materials Science and Engineering: A}, vol.~814, p.~141271, 2021.

\bibitem{bkm:RefKelly2004-102}
S.~Kelly, {\em Thermal and microstructure modeling of metal deposition
  processes with application to \uppercase{T}i-6\uppercase{A}l-4\uppercase{V}}.
\newblock Phd thesis, Virginia Tech, 2004.

\bibitem{bkm:RefIdhil2016-103}
A.~I. Ismail, M.~Dehmas, E.~Aeby-Gautier, and B.~Appolaire, ``In-situ
  investigation of phase transformation kinetics in
  \uppercase{T}i-6\uppercase{A}l-4\uppercase{V} under rapid heating condition
  using high-energy synchrotron diffraction,'' {\em Proceedings of the 13th
  World Conference on Titanium}, pp.~591--598, 2016.

\bibitem{bkm:RefHooper2018-94}
P.~Hooper, ``Melt pool temperature and cooling rates in laser powder bed
  fusion,'' {\em Additive Manufacturing}, vol.~22, pp.~548--559, 2018.

\bibitem{bkm:RefAhmed1998-95}
T.~Ahmed and H.~Rack, ``Phase transformations during cooling in $\alpha+\beta$
  titanium alloys,'' {\em Materials Science and Engineering: A}, vol.~243,
  pp.~206--211, 1998.

\bibitem{bkm:RefCallister2011-96}
W.~Callister and D.~Rethwisch, {\em Materials science and engineering}.
\newblock NJ, USA: Wiley: Hoboken, 2011.

\bibitem{bkm:RefBrown2021-97}
D.~Brown, V.~Anghel, L.~Balogh, B.~Clausen, N.~S. Johnson, R.~M. Martinez,
  D.~C. Pagan, G.~Rafailov, L.~Ravkov, M.~Strantza, and E.~Zepeda-Alarcon,
  ``Evolution of the microstructure of laser powder bed fusion
  \uppercase{T}i-6\uppercase{A}l-4\uppercase{V} during post-build heat
  treatment,'' {\em Metallurgical and Materials Transactions A}, vol.~52,
  pp.~5165--5181, 2021.

\bibitem{bkm:RefQazi2003-98}
J.~Qazi, O.~Senkov, J.~Rahim, and F.~Froes, ``Kinetics of martensite
  decomposition in \uppercase{T}i-6\uppercase{A}l-4\uppercase{V}-x\uppercase{H}
  alloys,'' {\em Materials Science and Engineering: A}, vol.~359, pp.~137--149,
  2003.

\bibitem{bkm:RefNeelakantan2009-99}
S.~Neelakantan, P.~R.-D. del Castillo, and S.~van~der Zwaag, ``Prediction of
  the martensite start temperature for $\beta$ titanium alloys as a function of
  composition,'' {\em Scripta Materialia}, vol.~60, pp.~611--614, 2009.

\bibitem{bkm:RefChong2017-100}
Y.~Chong, T.~Bhattacharjee, J.~Yi, A.~Shibata, and N.~Tsuji, ``Mechanical
  properties of fully martensite microstructure in
  \uppercase{T}i-6\uppercase{A}l-4\uppercase{V} alloy transformed from refined
  beta grains obtained by rapid heat treatment \uppercase{(RHT)},'' {\em
  Scripta Materialia}, vol.~138, pp.~66--70, 2017.

\bibitem{bkm:RefIvasishin1999-101}
O.~Ivasishin and R.~Teliovich, ``Potential of rapid heat treatment of titanium
  alloys and steels,'' {\em Materials Science and Engineering: A}, vol.~263,
  pp.~142--154, 1999.

\bibitem{semiatin2005prediction}
S.~Semiatin, N.~Stefansson, and R.~Doherty, ``Prediction of the kinetics of
  static globularization of ti-6al-4v,'' {\em Metallurgical and Materials
  Transactions A}, vol.~36, pp.~1372--1376, 2005.

\bibitem{stefansson2002kinetics}
N.~Stefansson, S.~Semiatin, and D.~Eylon, ``The kinetics of static
  globularization of ti-6al-4v,'' {\em Metallurgical and Materials Transactions
  A}, vol.~33, pp.~3527--3534, 2002.

\bibitem{granasy2019phase}
L.~Gr{\'a}n{\'a}sy, G.~I. T{\'o}th, J.~A. Warren, F.~Podmaniczky, G.~Tegze,
  L.~R{\'a}tkai, and T.~Pusztai, ``Phase-field modeling of crystal nucleation
  in undercooled liquids--a review,'' {\em Progress in Materials Science},
  vol.~106, p.~100569, 2019.

\bibitem{zhao2023influence}
R.~Zhao, X.~Yan, H.~Wang, C.~Song, C.~Li, L.~Mao, M.~Liu, J.~Gao, and Z.~Sun,
  ``Influence of non-equilibrium solidification of melt pools and annealing on
  microstructure formation and mechanical properties of laser powder bed
  fusion-built ti--6al--4v alloys,'' {\em Materials Science and Engineering:
  A}, vol.~873, p.~144964, 2023.

\bibitem{du2023facile}
X.~Du, M.~Simonelli, J.~W. Murray, and A.~T. Clare, ``Facile manipulation of
  mechanical properties of ti-6al-4v through composition tailoring in laser
  powder bed fusion,'' {\em Journal of Alloys and Compounds}, vol.~941,
  p.~169022, 2023.

\bibitem{mckenna2023evaluation}
T.~McKenna, C.~Tomonto, G.~Duggan, E.~Lalor, S.~O'Shaughnessy, and D.~Trimble,
  ``Evaluation of bimodal microstructures in selective-laser-melted and
  heat-treated ti-6al-4v,'' {\em Materials \& Design}, vol.~227, p.~111700,
  2023.

\bibitem{wang2022formation}
H.~Wang, Q.~Chao, H.~Chen, Z.~Chen, S.~Primig, W.~Xu, S.~Ringer, and X.~Liao,
  ``Formation of a transition v-rich structure during the $\alpha$'to $\alpha$+
  $\beta$ phase transformation process in additively manufactured ti-6al-4 v,''
  {\em Acta Materialia}, vol.~235, p.~118104, 2022.

\bibitem{bkm:RefLucarini2019-104}
S.~Lucarini and J.~Segurado, ``On the accuracy of spectral solvers for
  micromechanics based fatigue modeling,'' {\em Computational Mechanics},
  vol.~63, pp.~365--382, 2019.

\bibitem{xiang2023phase}
H.~Xiang, W.~Van~Paepegem, and L.~Kestens, ``Phase-field simulation of
  martensitic transformation with different conditions in inhomogeneous
  polycrystals,'' {\em Computational Materials Science}, vol.~220, p.~112067,
  2023.

\bibitem{liu2020integration}
P.~Liu, Z.~Wang, Y.~Xiao, R.~A. Lebensohn, Y.~Liu, M.~F. Horstemeyer, X.~Cui,
  and L.~Chen, ``Integration of phase-field model and crystal plasticity for
  the prediction of process-structure-property relation of additively
  manufactured metallic materials,'' {\em International Journal of Plasticity},
  vol.~128, p.~102670, 2020.

\end{thebibliography}

\end{document}